\documentclass{aa}
\usepackage{txfonts}
\usepackage{graphicx, aas_macros}
\usepackage[round]{natbib}
\bibpunct{(}{)}{;}{a}{}{,}
\usepackage[scriptsize]{subfigure}
\usepackage{dcolumn}

\begin{document}
\newcolumntype{.}{D{.}{.}{5}}
\newcolumntype{,}{D{~}{\,}{-1}}
   \title{Motion and Properties of Nuclear Radio Components in Seyfert Galaxies Seen with VLBI}

   \author{E. Middelberg,
          \inst{1}
          A. L. Roy,
          \inst{1,7}
          N. M. Nagar,
          \inst{2,9}
          T. P. Krichbaum,
          \inst{1}
          R. P. Norris,
          \inst{3}
          A. S. Wilson,
          \inst{4}
          H. Falcke,
          \inst{1,8}
          E. J. M. Colbert,
          \inst{5}
          A. Witzel
          \inst{1}
          \and
          K. J. Fricke,
          \inst{6}
          }

   \offprints{E. Middelberg}

   \institute{Max-Planck-Institut f\"ur Radioastronomie,
              Auf dem H\"ugel 69, D-53121 Bonn, Germany\\
              \email{emiddelb, aroy, tkrichbaum, awitzel@mpifr-bonn.mpg.de}
         \and
              Kapteyn Institute, University of Groningen,
              Landleven 12, 9747 AD Groningen, The Netherlands\\
              \email{nagar@astro.rug.nl}
         \and
              Australia Telescope National Facility,
              PO Box 76, Epping, NSW, Australia\\
              \email{rnorris@atnf.csiro.au}
         \and
              Department of Astronomy,
              University of Maryland,
              College Park, MD 20742, USA\\
              \email{wilson@astro.umd.edu}
         \and
              John Hopkins University,
              Department of Physics and Astronomy,
              3400 North Charles Street,
              Baltimore, MD 21218, USA\\
              \email{colbert@pha.jhu.edu}
         \and
              Universit\"ats-Sternwarte,
              Geismarlandstrasse 11,
              D-37083 G\"ottingen, Germany\\
              \email{kfricke@uni-sw.gwdg.de}
         \and
              Geod\"atisches Institut der Universit\"at Bonn,
              Nussallee 17, D-53115 Bonn, Germany
         \and
              ASTRON, P.O. Box 2,
              7990 AA Dwingeloo, The Netherlands\\
              \email{falcke@astron.nl}
         \and
              INAF,
              Arcetri Observatory, 
              Largo E. Fermi 5, Florence 50125, Italy\\
   \date{Received January 1, 2002; accepted January 1, 2004}
            }

\abstract{
We report EVN, MERLIN and VLBA observations at 18\,cm, 6\,cm and
3.6\,cm of the Seyfert galaxies \object{NGC~7674}, \object{NGC~5506},
\object{NGC~2110} and \object{Mrk~1210} to study their structure and
proper motions on pc scales and to add some constraints on the many
possible causes of the radio-quietness of Seyferts. The component
configurations in \object{NGC~7674} and \object{NGC~2110} are simple,
linear structures, whereas the configurations in \object{NGC~5506} and
\object{Mrk~1210} have multiple components with no clear axis of
symmetry. We suggest that \object{NGC~7674} is a low-luminosity
compact symmetric object. Comparing the images at different epochs, we
find a proper motion in \object{NGC~7674} of $(0.92\pm0.07)~c$ between
the two central components separated by 282~pc and, in
\object{NGC~5506}, we find a $3~\sigma$ upper limit of $0.50~c$ for
the components separated by $3.8~{\rm pc}$. Our results confirm and
extend earlier work showing that the outward motion of radio
components in Seyfert galaxies is non-relativistic on pc scales. We
briefly discuss whether this non-relativistic motion is intrinsic to
the jet-formation process or results from deceleration of an initially
relativistic jet by interaction with the pc or sub-pc scale
interstellar medium.  We combined our sample with a list compiled from
the literature of VLBI observations made of Seyfert galaxies, and
found that most Seyfert nuclei have at least one flat-spectrum
component on the VLBI scale, which was not seen in the spectral
indices measured at arcsec resolution. We found also that the bimodal
alignment of pc and kpc radio structures displayed by radio galaxies
and quasars is not displayed by this sample of Seyferts, which shows a
uniform distribution of misalignment between $0^{\circ}$ and
$90^{\circ}$. The frequent misalignment could result from jet
precession or from deflection of the jet by interaction with gas in
the interstellar medium.\\
\keywords{Galaxies: Seyfert, Galaxies: active, Galaxies: jets} }

\authorrunning{Middelberg et al.}
\titlerunning{Motion and Properties of Nuclear Radio Components in Seyfert Galaxies Seen with VLBI}

   \maketitle

\section{Introduction}

Active galaxies are conventionally divided into radio-quiet and
radio-loud categories. \cite{1989AJ.....98.1195K} found that quasars
separated into these two classes when the ratio of their radio to
optical continuum was considered, and a similar result has been found
by most (e.g. \citealt{1986MNRAS.218..265P},
\citealt{1990MNRAS.244..207M},
\citealt{2002AJ....124.2364I}) but not all (\citealt{2000ApJS..126..133W})
workers since. Members of the radio-loud category include BL~Lac
objects, radio galaxies and radio-loud quasars. Radio-loud objects
have large-scale radio lobes and compact luminous cores which often
exhibit apparent superluminal motion. Radio-quiet objects include
Seyfert galaxies and radio-quiet quasars; most of their radio emission
is confined to the sub-kpc scale.

Recent Very Long Baseline Interferometry (VLBI) has shown that many
Seyferts also contain compact nuclear radio components with brightness
temperatures $\gg10^7~{\rm K}$ despite having radio luminosities over
a factor of $10^8$ less than the most powerful active galaxies
(e.g. \citealt{1998ApJ...496..196U}; \citealt{1999ApJ...516..127U};
\citealt{1999ApJ...517L..81U}; \citealt{2000evn..proc....7R};
\citealt{2000ApJ...529..816M}), and that they harbour compact
components moving at non-relativistic speeds. In all cases, the
central masses are inferred to be black holes of millions to billions
of solar masses, so the central potentials are relativistically
deep. Since most accretion energy is released close to the hole, jets
might be expected to be relativistic in all cases. Why then are the
radio sources in Seyferts small, weak and slow compared to those in
radio-loud objects?

Various answers to this question have been proposed, involving either
``intrinsic'' differences in the central engine or ``extrinsic''
differences in the surrounding medium. Intrinsic differences that have
been suggested include (i) systematically lower black hole masses in
Seyferts (\citealt{2000ApJ...543L.111L}), (ii) lower black hole spins
(\citealt{1995ApJ...438...62W}), (iii) a ``magnetic switch'' that was
identified by \cite{1997Natur.388..350M} during numerical modeling of
jets, (iv) the production in Seyferts of buoyant plasmons that bubble
up through the density gradient of the narrow line region (NLR)
instead of a collimated relativistic jet
(\citealt{1985MNRAS.214..463P}, \citealt{1986MNRAS.222..189W},
\citealt{1989MNRAS.240..487T}), (v) a large thermal plasma fraction
in the jet in Seyferts (\citealt{1998ApJ...495..680B}), or (vi)
radiative inefficiency (\citealt{Falcke1995}). Extrinsic differences
generally invoke the rapid deceleration of initially relativistic jets
by collisions in a dense surrounding BLR or ISM
(e.g. \citealt{1984AaA...141...85N}). With so many possible causes,
and so few observational constraints, the question remains open why
Seyfert galaxies have such low radio luminosities.

One approach to investigate whether the radio quietness of Seyferts is
due to intrinsic or extrinsic causes is to measure speeds of radio
components as close to the nucleus as possible, hopefully before
interactions with the ISM affect them much. How close that needs to be
is unclear, but signs of jet-NLR interaction have been seen on scales
of tens of pc (\citealt{1986MNRAS.222..189W};
\citealt{1994AJ....107.1227W};
\citealt{1995ApJ...454..106B}; \citealt{1997ApJ...487..560C};
\citealt{1998ApJ...502..199F}), so sub-pc scales are desirable. Proper
motions of Seyfert galaxy radio components have now been measured with
VLBI in a few cases (e.g. \object{Mrk~231} and \object{Mrk~348} by
\citealt{1999ApJ...517L..81U}, \object{NGC~1068} by
\citealt{2000evn..proc....7R}, and \object{NGC~4151} by
\citealt{1998ApJ...496..196U}, \citealt{2002AAS...201.4814U}), and the
motions were found to be $\le 0.25~c$. In \object{III~Zw~2}, $v<0.04~c$ was
seen between two barely-resolved components over eight months,
followed by a sudden displacement at $v\ge1.25~c$ over seven months
(\citealt{2000AaA...357L..45B}).

The observations presented here were made for three reasons: 1) to
measure proper motions of radio components on pc scales.  2) to look
for extended, low surface-brightness radio emission on the scale of
the NLR.  In the event that slow proper motions were measured, the
next question would be whether components were ejected slowly or
whether they were ejcted at high, perhaps relativistic velocities and
then slowed by interaction with the NLR gas.  Were such braking to
occur, one would expect shocks in the jet and possibly jet disruption.
The in-situ particle acceleration associated with shocks could lead
then to extended regions of radio emission.  3) to measure spectral
indices, which provide a useful diagnostic of emission mechanisms,
source compactness, and possible foreground free-free absorption.
Most spectral index measurements of Seyfert galaxies to date have used
the VLA, ATCA or WSRT, whose relatively large beams can average
together many components in the nucleus. Measurements with
parsec-scale resolution in Seyfert galaxies enable one to separate a
possible compact, flat-spectrum component from surrounding
steep-spectrum, optically-thin synchrotron emission, but such
measurements require VLBI with matched beams at two frequencies and
are still relatively few. In contrast, powerful radio sources have
been well studied with VLBI and flat-spectrum core components are
often seen at the base of a conically-expanding jet, thus indicating
the location of the nucleus.  Do Seyfert nuclei also show a single,
compact, synchrotron self-absorbed core component at the base of a
collimated jet?\\

To meet these goals, we observed using Global VLBI, the European VLBI
Network (EVN), the Multi-Element Radio Linked Interferometer Network
(MERLIN) and the Very Long Baseline Array (VLBA) the Seyfert galaxies
\object{NGC~5506} and \object{NGC~7674} at multiple epochs for proper motion
measurements of nuclear radio sources, and \object{Mrk~1210} and
\object{NGC~2110} for the first epoch.  To enlarge the sample size for
deriving statistics on the spectral and geometric properties of the pc
scale structure of Seyferts, we made a complete literature review of
VLBI observations of Seyfert galaxies. We assume $H_0 = 75~{\rm
km\,s^{-1}\,Mpc^{-1}}$ and $q_0$ = 0.5 throughout this paper, and
calculate spectral indices using $S\propto\nu^{\alpha}$.

\section{The Sample and Observations}\label{sec:observations}

The galaxies presented here are part of a sample selected for global
VLBI observations in 1994 to investigate the dichotomy between
radio-loud and radio-quiet AGN. The sample comprised 16 Seyfert 2
galaxies from \cite{1991cqan.book.....V} with 6~cm core flux densities
$>70~{\rm mJy}$ and $\delta>-10^{\circ}$. The literature was then
carefully reviewed for updated spectral classifications, some of the
objects were discarded and others were incorporated based on their new
classification and flux density measurements.

{\bf Global VLBI, 1994.}~~The sample was observed in 1994 with global
VLBI at 6~cm, using the ten VLBA stations, the phased VLA, Effelsberg,
Onsala, Medicina and Noto for 36~min in six snapshots over 11~h,
yielding a $\sim30~{\rm min}$ integration of each object (Krichbaum et
al. 1994, unpublished). The observations yielded only the detection of
point sources in the cases of \object{NGC~7674}, \object{NGC~2110} and
\object{Mrk~1210}, and lacked short baselines and spectral index
information. Only the \object{NGC~5506} data are included in this
paper.

{\bf EVN and MERLIN, 1999/2000.}~~Four of those objects
(\object{NGC~5506}, \object{NGC~7674}, \object{NGC~2110} and
\object{Mrk~1210}) were re-observed in 1999 and 2000 with better
sensitivity and better $(u,v)$-plane coverage using EVN and MERLIN at
18~cm and 6~cm.

\begin{table*}
\begin{center}
\begin{tabular}{l..ccll}
\hline
\hline
Source & 
\multicolumn{1}{l}{\rm RA (J2000)} & 
\multicolumn{1}{l}{\rm DEC (J2000)} & 
$D$ &
Type &
Purpose & 
{\rm Position reference}\\
\hline
\object{NGC~5506} & 14~13~14.87926 & -03~12~27.6514  & 24.7$^{1a)}$ & S1n$^{1b)}$ & Science Target   & Roy et al. (1997)$^{1c)}$\\
\object{1404-015} & 14~04~45.8949  & -01~30~21.937   &              &             & Phase calibrator & -\\
\object{NGC~7674} & 23~27~56.712   & +08~46~44.14    & 116$^{2a)}$  & S1h$^{2b)}$ & Science Target   & Krichbaum et al. (1994)$^{2c)}$\\
\object{2327+096} & 23~27~33.5808  & +09~40~09.460   &              &             & Phase calibrator & \cite{1998MNRAS.293..257B}\\
\object{NGC~2110} & 05~52~11.376   & -07~27~22.52    & 31.1$^{3a)}$ & S1i$^{3b)}$ & Science Target   & \cite{1983ApJ...264L...7U}$^{3c)}$\\
\object{0541-056} & 05~41~38.0848  & -05~41~49.395   &              &             & Phase calibrator & -\\
\object{Mrk~1210} & 08~04~05.856   & +05~06~49.83    & 53.9$^{4a)}$ & S1h$^{4b)}$ & Science Target   & \cite{1998ApJ...502..199F}$^{4c)}$\\
\object{0803+043} & 08~03~56.4444  & +04~21~02.724   &              &             & Phase calibrator & \cite{1998MNRAS.293..257B}\\
\hline
\end{tabular}
\caption[List of observed sources]{List of observed sources. Column 2
and 3 give the coordinates used in the observations and for
correlation, column 4 gives the source distances in Mpc. Column 5
gives the classification according to \cite{Veron2001}. Remarks:
1a) \cite{1996ApJS..106...27K},
1b) \cite{2002AaA...391L..21N}, 
1c) Coordinates from Roy et al. (1997) observations (unpublished), which based on VLA observations by \cite{1984ApJ...285..439U},
2a) \cite{2000AJ....120.1691N}, 
2b) \cite{Veron2001}, 
2c) VLA 6 cm (unpublished), 23 25 24.410 +08 30 12.60 (B1950) precessed to J2000 by NRAO's scheduling software SCHED, \cite{1996ApJS..106...27K}, 
3a) \cite{1995ApJS...99...67N}, 
3b) \cite{1998csan.book.....V}, 
3c) 05~49~46.376 $-$07~28~01.99 (B1950) precessed to J2000 by SCHED, 
4a) \cite{1991trcb.book.....D}, 
4b) \cite{1998csan.book.....V}, 
4c) Based on 6 cm VLA observations from \cite{1998ApJ...502..199F}.}
\label{tab:sourcelist}
\end{center}
\end{table*}

The observations were made with eight EVN stations (Effelsberg,
Cambridge 32~m, Jodrell Bank Mk~1 at 18~cm and Jodrell Bank Mk~2 at
6~cm, Medicina, Noto, Onsala 25~m, Westerbork array, Torun) at 18~cm
(1.65\,GHz) on November 11, 1999, and at 6~cm (4.99\,GHz) on February
28, 2000. The 18~cm EVN observations were recorded in MkIII mode with
a bandwidth of 56~MHz, 1 bit sampling and dual circular polarization,
and were correlated in Bonn on the MkIII correlator. The 6~cm EVN
observations were recorded in the new MkIV mode with a bandwidth of
32~MHz, 2 bit sampling and dual circular polarization, and were
correlated at the Joint Institute for VLBI in Europe (JIVE). The
MERLIN observations were correlated in real time with a bandwidth of
32~MHz, 2 bit sampling and dual circular polarization. As all sources
are weak, phase-referencing was used for initial phase
calibration. Details of the calibrators are given in Table
\ref{tab:sourcelist}. All calibrators lie within $6^{\circ}$ of the
targets. Phase calibrator scans were obtained every 5~min to 10~min at
18~cm and every 4~min at 6~cm.

{\bf Additional data.}~~Further observations presented in
this paper were done in the framework of other projects:

\begin{itemize}

\item \object{NGC~5506} was observed during two further VLBI experiments,
designed for proper motion measurement.

(i) February 5, 1997: \object{NGC~5506} was observed with the ten VLBA
stations at 18~cm, 6~cm and 3.6~cm with on-source times of 91~min,
64~min and 43~min, a bandwidth of 32~MHz, 2 bit sampling and right
circular polarization.

(ii) May 31, 2000: \object{NGC~5506} was observed with the ten VLBA
stations at 18~cm and 6~cm with on-source times of 204~min at each
wavelength, a bandwidth of 32~MHz, 2 bit sampling and left circular
polarization.

The 1994 and 2000 VLBA observations and the EVN observations were
reduced in AIPS using the same procedure for each (see below),
including the same weighting, taper and restoring beam size, and the
1997 VLBA observations were reduced using the same methods except for
slightly different tapering. All 18~cm and 6~cm images are shown in
Fig.~\ref{fig:ngc5506_motion}.

\item \object{NGC~7674} was observed at 18~cm on April 13, 1985 by
\cite{1988MNRAS.234..745U} with four EVN stations using a bandwidth of
28~MHz and dual circular polarization. We re-reduced those data from
the archive in Bonn following the same procedures as for the 1999 EVN
18~cm data, yielding images with ${\rm rms=0.46~mJy\,beam^{-1}}$ and a
beam size of $39.3\times24.2~{\rm mas}$. Amplitude calibration data
were not archived, so we used $T_{\rm sys}$ and gain measurements from
the immediately preceding experiment (Porcas, priv. comm.).

\item \object{Mrk~1210} was observed on August 6, 1998 with the
VLBA at 18 and 6~cm with on-source times of 160~min and 140~min, a
bandwidth of 32~MHz, 2 bit sampling and dual circular
polarization. Phase-referencing was made to the same calibrator as for
the EVN and MERLIN observations. The observations yielded thermal
noise limited images with resolutions of $18.6\times14.2~{\rm mas}$
and $3.9\times1.8~{\rm mas}$ at 18~cm and 6~cm.\\

\end{itemize}

Data reduction was done using the {\it A}stronomical {\it I}mage {\it
P}rocessing {\it S}ystem (AIPS). The EVN and VLBA amplitudes were
calibrated using $T_{\rm sys}$ measurements and the MERLIN amplitudes
were calibrated using the flux density calibrator \object{DA~193}
assuming flux densities of 2.12~Jy at 18~cm and 5.15~Jy at
6~cm. Residual delays and phases were determined by fringe fitting on
the phase reference calibrators and interpolating the solutions onto
the targets. If the target was visible in the resulting
phase-referenced image, one to three iterations of phase
self-calibration and deconvolution using CLEAN were performed until a
thermal noise limited image was achieved. Absolute position
information is then preserved as the self-calibration uses
phase-referenced, and therefore position-referenced, model
components. If the target was not visible after phase referencing, the
data were self-calibrated using a point source as the initial
model. In this case, absolute position information was lost (see
Table~\ref{tab:results}).

Absolute source positions were measured from the phase-referenced,
self-calibrated images, where available; otherwise, relative source
positions were measured from the self-calibrated images. Positions
were always measured by fitting a two-dimensional parabola to the
source and recording the position of the peak. This was found superior
to fitting Gaussians because the images of the sources were not always
well described by a Gaussian. Phase-referenced positions are based on
the calibrator positions in Table~\ref{tab:sourcelist}. Absolute
position uncertainties are typically $10^{-7}$ times the
source-calibrator separation (\citealt{1991ritt.proc..289L}). Our
random position uncertainties should be a small fraction of the
beamwidth, and proportional to beamwidth/SNR, where SNR is the signal
to noise ratio of the source component.

Flux densities were measured by integrating over the source region in
the image plane. The estimated $1~\sigma$ flux density errors comprise
a $5~\%$ (MERLIN) or $10~\%$ (other arrays) flux density scale
calibration uncertainty, $\approx0.2~{\rm mJy}$ additive thermal
noise, the exact value depending on the source size and the particular
image, and $5~\%$ to $10~\%$ uncertainty in the integration over the
source region, depending on the size and flux density. The combined
estimated errors of the integrated flux densities in
Table~\ref{tab:results} represent image rms noise plus $11~\%$
($S_{\nu}<20~{\rm mJy}$) and $7~\%$ ($S_{\nu}>20~{\rm mJy}$) for
MERLIN observations and $15~\%$ ($S_{\nu}<20~{\rm mJy}$) and $11~\%$
($S_{\nu}>20~{\rm mJy}$) for other VLBI observations. The peak flux
density errors are image rms noise plus $5~\%$ for MERLIN observations
and $10~\%$ for other VLBI observations.

Spectral indices were measured by tapering the 6~cm data and restoring
with the 18~cm beam size. The uncertainty on the spectral indices,
based on the uncertainties on the flux density measurements is 0.09
($1~\sigma$) for a 20~mJy component. Despite the taper, some beam
mismatch remained because the 18~cm and tapered 6~cm arrays were not
perfectly scaled arrays; the resulting magnitude of the effect is
difficult to quantify because it depends on the source structure. The
largest effect probably comes from the short baselines at 18~cm, since
these were not removed to match the 6~cm short-baseline length. This
would affect sources that have emission that is extended on scales
larger than the largest angular scale to which the 6~cm observations
are sensitive (11~mas for the EVN), but still small enough to be
detected at 18~cm ($<35~{\rm mas}$ for the EVN). The insensitivity of
the 6~cm observations to emission on this range of scales makes the
spectral index appear more negative than it actually is. We see
emission extended on this scale in the case of \object{NGC~5506}
(components B0 and B1) and \object{NGC~7674} (\citealt{Momjian2003}),
and we attempt in these cases to estimate a correction. For the other
sources our EVN 18~cm images and MERLIN images do not show structure
extended on this scale and so we believe that no bias in the spectral
indices should be present.

\section{Results and Discussion}\label{sec:results}

\subsection{\object{NGC~5506}}

\object{NGC~5506} is an edge-on, dusty irregular or early-type spiral. It has
formerly been classified as a type 2 Seyfert based on the nuclear
optical spectrum (\citealt{1976MNRAS.177..673W}), but recent near-IR
spectra suggest that it is a narrow-line Seyfert 1
(\citealt{2002AaA...391L..21N}).

\cite{1996ApJ...467..551C} detected large-scale radio emission extending up
to 3.6 kpc away from the disk in a north-south direction
(i.e. perpendicular to the disk), but could not establish whether this
is starburst or AGN driven. At higher resolution
($\sim0.7^{\prime\prime}$), a MERLIN 18~cm image by
\cite{1986MNRAS.219..387U} shows that \object{NGC~5506} has a compact core
with a diffuse halo of a few arcseconds diameter. The compact core is
slightly extended and aligns with the major axis of the host galaxy
and with the position angle of the optical continuum polarization
(\citealt{1983ApJ...266..470M}). This is contrary to the situation in
most Seyfert 2s, in which the polarization tends to be perpendicular
to the jet axis (\citealt{Antonucci1985}) and strengthens the Sy~1
classification.

A water maser was discovered in \object{NGC~5506} by \cite{1994ApJ...437L..99B}
during a survey of Seyfert galaxies. Unfortunately, its weakness,
60~mJy in the broad component, has prevented detailed VLBI imaging.\\

\subsubsection{Results}

\begin{figure}[ht!]
\includegraphics[width=\linewidth]{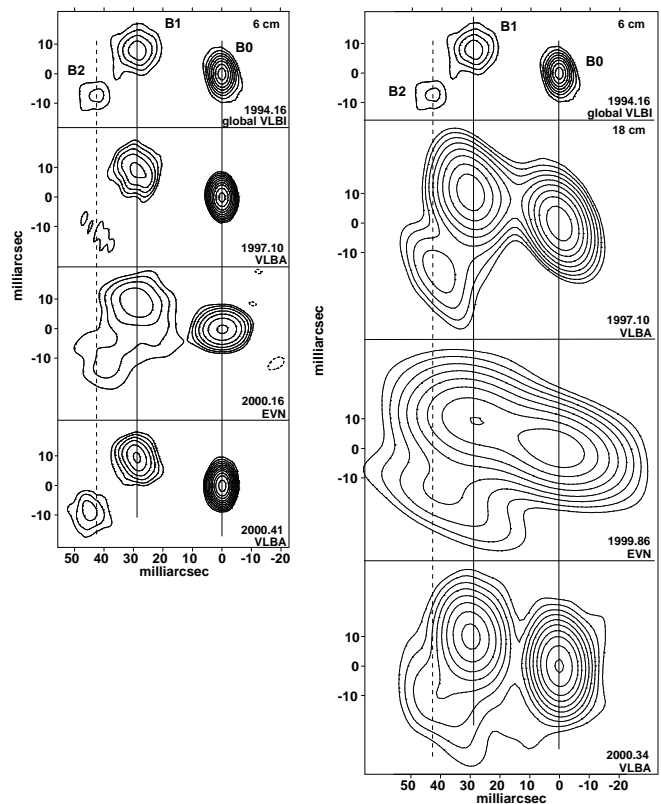}
  \caption{VLBI 6~cm images at four epochs (left panels), from which
  we placed an upper limit on the proper motion of the B1 component,
  relative to B0, in \object{NGC~5506}. The proper motion of the B2 component
  is unreliable due to limited image fidelity.  The 6~cm 1994.16 epoch
  is repeated as the top right panel. The lower three right panels
  show the 18~cm images. All images are naturally weighted. The VLBA
  observations were restored with common beams of $7.4\times4.3~{\rm
  mas}$ (6~cm) and $18.5\times11.3~{\rm mas}$ (18~cm).}
\label{fig:ngc5506_motion}
\end{figure}

\begin{figure}
  \includegraphics[width=\linewidth]{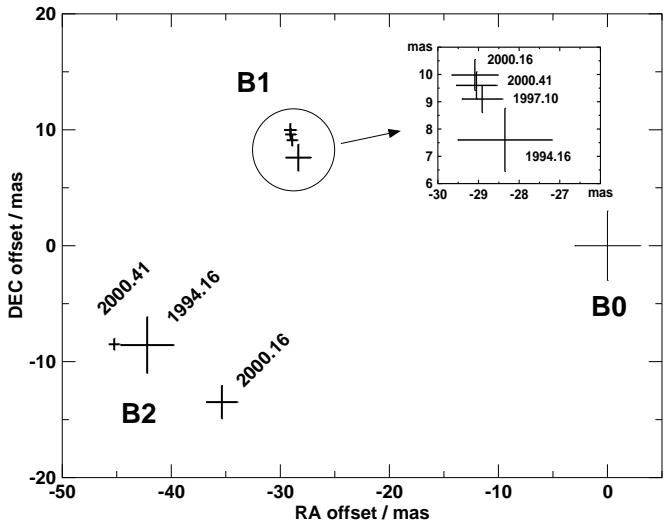}
  \caption{Diagram illustrating the apparent motions in \object{NGC~5506}. The
  peak positions of B1 and B2 from 6~cm global VLBI, EVN and VLBA are
  plotted relative to B0. Errors are $1~\sigma$.}
  \label{fig:vector_diagram}
\end{figure}

Both the MERLIN 18~cm and 6~cm images (not shown) display a single
unresolved point source with a total flux density of 113~mJy and
86~mJy, respectively. The single-dish 6~cm flux density in a
$2.4^{\prime}$ beam measured simultaneously at Effelsberg was 193~mJy,
so there is significant emission on scales larger than the MERLIN 6~cm
beam size, almost all of which appears in the VLA image of
\cite{1996ApJ...467..551C}.

The VLBI 18~cm and 6~cm images in Fig.~\ref{fig:ngc5506_motion} show
three components which we name B0 through B2, with separations of
3.5~pc (B0-B1) and 5.1~pc (B0-B2). These three components are
unresolved or slightly resolved by the VLBA and EVN. The total
18~cm flux density recovered by the EVN in 1999.86 was 65~mJy, one
third of the 6~cm single-dish emission.

We have aligned images from all four epochs using the center of B0 as
a reference point to illustrate the relative motions between
components inside \object{NGC~5506} (Fig.~\ref{fig:vector_diagram}). From
1994.16 to 1997.10, component B1 has moved $(1.60\pm1.26)~{\rm mas}$
($1~\sigma$ error), yielding $(0.21\pm0.17)~c$. However, the error is
large, and motion as fast as $0.50~c$ would be required for a
$3~\sigma$ detection. After 1997.10, the motion of B1 is
uncertain. The position angle in which B1 might be moving is
$(18\pm2)^{\circ}$, different from the P.A. of $(72\pm2)^{\circ}$ of
B1 with respect to B0, so if the motion of B1 is real, it is not
simply radially away from B0. Component B2 showed no consistent
believable displacements; the apparent motions of B2 are probably
due to limited image fidelity at low SNR.

The diameters of components B0 and B1 were measured by fitting a
Gaussian to the images, and the intrinsic size was determined by
deconvolving the beam from the fitted Gaussian. Component B0 proved to
be only slightly extended ($3.2\times2.6~{\rm mas}$) at 6~cm compared
to the beam size of $5.7\times3.7~{\rm mas}$ (VLBA, February 1997),
consistent with the near equality of peak and integrated flux
densities.  Component B1 was significantly resolved in the same
observation ($12.5\times8.9~{\rm mas}$), confirmed by the
significantly smaller peak flux density ($4.3~{\rm mJy\,beam^{-1}}$)
compared to the integrated flux density (24.4~mJy). The brightness
temperature of component B0, derived from the integrated flux density
and deconvolved component size, is $3.6\times10^8~{\rm K}$, whereas
the brightness temperature of the resolved component B1, derived from
the peak flux density and beam size, is $1.4\times10^7~{\rm K}$.

Spectral indices were measured using EVN 18~cm and EVN 6~cm images
tapered to matching resolution. The strongest component (B0) was found
to have a flat spectrum ($\alpha=-0.03$) while the other components
had steeper spectra ($\alpha=-0.77$ and $-0.3$).

\begin{figure}[ht!]
  \includegraphics[width=\linewidth]{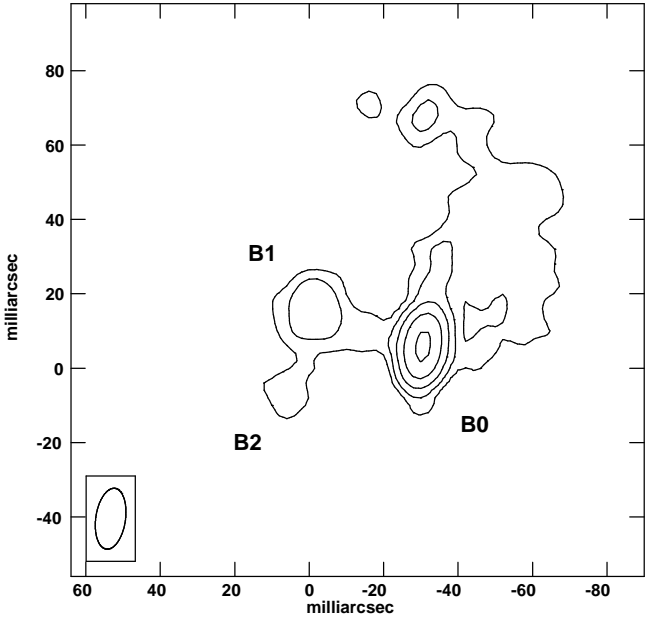}
  \caption{Combined EVN and MERLIN 6~cm image of \object{NGC~5506}. Offsets are
  relative to RA 14:13:14.87953, DEC $-$03:12:27.6799 (J2000), peak flux
  density is 37.3~mJy/beam, contours are
  1~mJy/beam$\times(2,4,8,16,32)$, beam size is 16.5$\times$8.0~mas
  and natural weighting was used.}
  \label{fig:ngc5506_e+m6}
\end{figure}

The spectral index measured for component B0 might be affected by the
extended emission to its NW (Fig~\ref{fig:ngc5506_e+m6}, adding a
negative bias to our measurement. To estimate the size of the effect,
we note that the extended emission has a brightness of $2~{\rm
mJy\,beam^{-1}}$ at 6~cm in a $16.5\times8.0~{\rm mas}$ beam, and is
present in one quadrant from B0. This emission would contribute a flux
density of 3~mJy in the $26.2\times14.1~{\rm mas}$ tapered beam area
at 18~cm assuming it has a spectral index of $-0.7$ and that it would
be seen by the EVN 18~cm array but not by the 6~cm array. Subtracting
3~mJy from the measured flux density of B0 at 18~cm of 32.9~mJy yields
a corrected spectral index of $+0.06$, as listed in
Table~\ref{tab:results}.

Likewise, the spectral index estimated for component B1 might be
affected by a possible bridge of faint emission between B1 and B0
visible in the combined EVN and MERLIN image in
Fig.~\ref{fig:ngc5506_e+m6}.  The emission has a brightness of $2~{\rm
mJy\,beam^{-1}}$ at 6~cm and occupies one octant.  This emission would
contribute 1.5~mJy to the EVN 18~cm measurement and would be resolved
out by the 6~cm EVN array.  Subtracting these 1.5~mJy from the
measured flux density of B1 at 18 cm of 25.6~mJy yields a corrected
spectral index of $-0.73$ (Table~\ref{tab:results}).

The flux density of component B0 remained constant to within the
uncertainties. Component B1 faded by a factor of two at 6~cm and
changed its spectral index from $\alpha^{18}_{6}=0.0$ in 1997 to
$\alpha^{18}_6=-0.77$ in 2000. The change could be due to an expanding
synchrotron source that was initially compact enough to be synchrotron
self-absorbed, or else could be due to changing foreground free-free
absorption.

Combining the EVN and MERLIN 6~cm data yields a possible bridge of
emission connecting the two brightest components and the detection of
very faint diffuse structure north of component B0
(Fig.~\ref{fig:ngc5506_e+m6}). This diffuse structure is interesting
because it suggests a physical connection between the AGN and the
radio emission on kpc scales observed with the VLA in C-configuration
by \cite{1996ApJ...467..551C}. The MERLIN 18~cm beam size of
$260\times 150~{\rm mas}$ includes the three components B0-B2 and the
diffuse northern emission ($80\times 40~{\rm mas}$) within a single
beam. Thus, the total MERLIN 18~cm flux of 113~mJy is probably the
combined emission from all the structures seen at 6~cm. This diffuse
northern emission is not seen in the EVN image because the EVN is
insensitive to structures larger than $\approx35~{\rm mas}$ at 18~cm
or 11~mas at 6~cm. Although component B2 has the same brightness as
the northern diffuse region in the MERLIN image, B2 remains visible in
the EVN images whereas the northern diffuse emission does not, because
B2 is more compact and so is not resolved away at the higher EVN
resolution.

\subsubsection{Discussion}
\label{sec:ngc5506_discussion}

The nuclear radio emission of \object{NGC~5506} is resolved into three
compact, non-colinear core components within the central 5~pc, and a
diffuse region of 8~pc extent to the north.  The two strongest radio
components (B0 and B1) are misaligned by $20^{\circ}$ from the
galactic plane and almost perpendicular to the kpc radio structure. On
the hundred pc to kpc scale, the radio jet axes in Seyferts have been
found to be randomly oriented with respect to the galactic disk
rotation axes (e.g. \citealt{1999ApJ...526L...9P},
\citealt{1999ApJ...516...97N},
\citealt{2000ApJ...537..152K}), whereas by conservation of angular
momentum during accretion, one would expect a common
axis. Misalignment of pc and kpc-scale radio emission is shown later
to be a common phenomenon in Seyfert galaxies
(Table~\ref{tab:misalignment}, Fig.~\ref{fig:histogram_PA}).\\

The radio emission from component B1 is peculiar. Its 6~cm integrated
flux density decreased from 30.0~mJy in 1994 to 24.4~mJy in 1997 and
10.8~mJy in February 2000, and then increased slightly to 15.3~mJy in
May 2000 (this last increase, however, is not significant within
$3\,\sigma$). Thus, the 6~cm flux density has constantly decreased,
producing a corresponding steepening of the spectrum over this six
year period from $\alpha=0.0$ in February 1997 to $\alpha=-0.78$ in
May 2000. The combination of flat spectrum and low brightness
temperature in 1997 ($1.4\times10^7~{\rm K}$) cannot be explained by
synchrotron self absorption, and is probably due to free-free
absorption. The time variability of the spectrum rules out free-free
emission because the spectral cutoff of a gas with such a high
temperature is in the X-ray regime and radio spectral index
variability as the component cools is therefore not expected.

B0 is a candidate for the location of the AGN core due to its flat
spectrum ($\alpha^{18}_6=0.06$) and compactness.

The alternative picture is that the AGN might lie between B0 and B1
and then the two bright components might be working surfaces where
double-sided jets from the AGN impact the ISM.  Low-surface-brightness
emission is seen extending from the hot spots to larger scales (the
region of emission north of B0, and component B2) and could be plasma
that has flowed through the hot spots and is either deflected by the
shock, or is rising buoyantly in the pressure gradient in the ISM.
However, B0 has a flat radio spectrum, indicative of optically thick
instead of optically thin synchrotron emission, whereas working
surfaces, like in compact symmetric objects (CSOs), characteristically
show steep spectra. Hence, we favour B0 being the core.

The picture of an AGN core lying between two hotspots seems to
describe the radio observations of \object{NGC~7674} discussed in the
next section and in \cite{Momjian2003}, and of NGC~3079 (Middelberg et
al., in prep.).

Ballistic motion of components, when back-extrapolated, should
converge on the core. However, in \object{NGC~5506} the B1 component,
if its motion is real, moves almost northwards in
P.A. $(18\pm2)^{\circ}$, whereas its position with respect to the
likely core component B0 is in P.A.  $(72\pm2)^{\circ}$.

A slight variant on this picture would have precession of the
accretion disk move the hot spots ``sideways'' along a ``wall'' of ISM
leaving behind a trail of fading extended plasma to account for the
observed proper motions having a tangential component, and for the
``trail'' of low surface-brightness emission behind each hot spot.

\subsection{\object{NGC~7674}}

NGC 7674 is a face-on spiral galaxy with asymmetric arms and a tidal
connection to a nearby compact elliptical galaxy
(\citealt{1996AJ....111..712H}). Observations with the VLA and EVN
(\citealt{1988MNRAS.234..745U}) reveal a linear triple radio source of
$\sim0.7^{\prime\prime}$ angular extent. The components are
$\sim0.5^{\prime\prime}$ west and $\sim0.15^{\prime\prime}$ east of
the main peak. \cite{Momjian2003} made a very sensitive observation of
\object{NGC~7674} with the VLBA, phased VLA and Arecibo at 1.4~GHz showing the
triple source found by \cite{1988MNRAS.234..745U} and additional low
surface-brightness emission forming an S-shaped structure.

\subsubsection{Results}

\begin{figure}[th!]
\centering
  \subfigure[\object{NGC~7674} at 18~cm (MERLIN). Offsets are relative to
  RA~23:27:56.71696, DEC~08:46:44.0078 (J2000), peak flux density is
  74.9~mJy/beam, contours are 1~mJy/beam$\times$($-5$, 5, 10, 15,
  20,~$\ldots$,~75 ), beam size is 276.1~mas$\times$110.4~mas and uniform
  weighting was used.]{\includegraphics[width=0.4\textwidth]{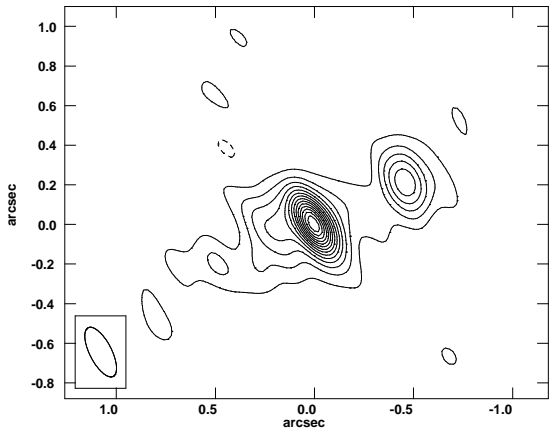}
  \label{fig:ngc7674_merlin_18cm}}\\
  \subfigure[EVN 18~cm images of \object{NGC~7674}.  
   {\it Top:} First epoch at 1985.28. Peak flux density is
  29.8~mJy/beam, contours are 1~mJy/beam$\times$($-2$, 2, 4, 8, 16, 25),
  beam size is 39.3~mas$\times$24.2~mas and natural weighting was
  used. {\it Bottom:} Second epoch at 1999.86. Offsets are relative to
  RA~23:27:56.71201, DEC~08:46:44:1368 (J2000), peak flux density is
  23.9~mJy/beam, contours are 1~mJy/beam$\times$($-1$, 1, 2, 4, 8, 16,
  32), beam size is 32.5~mas$\times$9.8~mas and uniform weighting was
  used.]{\includegraphics[width=0.35\textwidth,  angle=270]{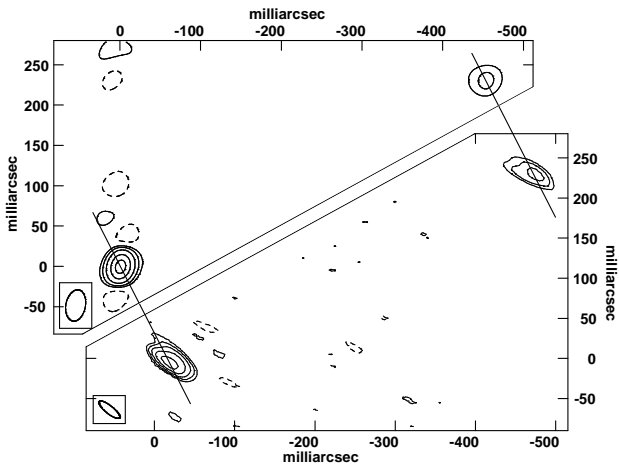}
  \label{fig:ngc7674_evn}}\\
  \caption{\object{NGC~7674} images}
\end{figure}

Our MERLIN 18~cm image (Fig.~\ref{fig:ngc7674_merlin_18cm}) shows the
well-known triple structure consistent with the VLA A-array 15~GHz
image by \cite{1988MNRAS.234..745U}, but the eastern component was not
detected by the EVN or by MERLIN at 6~cm.

The integrated MERLIN 6~cm flux density was 25.6~mJy, lower than the
total 6~cm single-dish flux density of 73~mJy in a $2.4^{\prime}$
beam. The latter was measured three months earlier at Effelsberg in a
related project to look for Seyfert variability. Thus, only $35~\%$ of
the total emission was detected by the MERLIN observation; the
remaining $65~\%$ is also not seen in the VLA 15~GHz image of
\cite{1988MNRAS.234..745U}. Both MERLIN and VLA observations are
insensitive to structures larger than 0.15~arcsec, or 90~pc.

Our 18~cm EVN image from November 11, 1999, in
Fig.~\ref{fig:ngc7674_evn} (lower panel) displays a bright, slightly
resolved component and point-like radio emission towards the north
west, consistent with the structure seen previously by
\cite{1988MNRAS.234..745U}.  The SE component is slightly extended in
the direction of the NW component. 

From the 1985 EVN archive observation we obtained a good-quality 18~cm
image (Fig.~\ref{fig:ngc7674_evn}, upper panel) with a $49~\sigma$
detection of the SE component and a $9~\sigma$ detection of the NW
component. We produced images from the 1985 and the 1999 data sets
with matching $(u,v)$ coverages and equal restoring beams
($41.9\times24.5\,{\rm mas^2}$ in P.A. $-29.20^{\circ}$) to look for a
change in separation of the components. From the 1985 image, we
measured a separation of $(508.7\pm0.4)~{\rm mas}$ ($1~\sigma$ error),
and from the 1999 image, we measured $(516.0\pm0.4)~{\rm mas}$
($1~\sigma$ error). Thus, we find that the relative separation between
the components has increased by $(7.3\pm0.6)~{\rm mas}$ in
14.58~years, corresponding to $0.28~{\rm pc\,yr^{-1}}$ or $v_{\rm
app}=(0.92\pm0.07)~c$. The separation increased along
P.A. $110^{\circ}$, in agreement with the P.A. of the the central and
the NW component of $117^{\circ}$. The 18~cm flux densities of the SE
and NW component have increased from 22.3~mJy to 37.6~mJy and from
4.1~mJy to 11.0~mJy, respectively, the second epoch being in agreement
with
\cite{Momjian2003}. Amplitude calibration errors could in principle
produce a scaling of the apparent flux densities, but probably not by
such a large factor as seen here. Furthermore, they cannot produce the
observed decrease of the ratio of the flux densities from 5.4 to
3.4. Thus, it is most likely that both components have brightened.

The spectral indices were calculated using the EVN 18~cm and EVN 6~cm
images tapered to matching resolution. The low surface brightness
emission surrounding component NW seen by \cite{Momjian2003}
($1.7~\mu{\rm Jy\,mas^2}$) has a negligibly small effect on the
spectral indices measured by us.

\subsubsection{Discussion}

All components in the EVN and MERLIN images of \object{NGC~7674} have
steep spectra, and no core identification can be made based on the
spectrum. \cite{Momjian2003} suggest that the nucleus might lie
between the two components we detected and is free-free absorbed at
21~cm. If so, our observations show that it must also be absorbed at
6~cm.

If the core is indeed located between the two brightest components,
then the structure would be reminiscent of that in compact symmetric
objects, and the components could be working surfaces where the jets
impact the ISM.  Then the proper motions would measure the hot-spot
advance speed.  From the proper motion of $0.14~{\rm pc\,yr^{-1}}$ and
the component separation of 282~pc we derive a dynamical age for the
source of 2000~yr, which lies within the range of ages measured for
CSOs by \cite{Polatidis2002}.  However, the 18~cm radio luminosity of
the two components in \object{NGC~7674} combined is $8\times 10^{22}~{\rm
W\,m^{-2}\,Hz^{-1}}$, which is three orders of magnitude less than the
typical CSOs studied by
\cite{Polatidis2002}.  The high radio luminosity of CSOs is probably a
selection effect, since the CSO samples are drawn from
flux-density-limited surveys and so automatically exclude
lower-luminosity sources.  Thus, we may here be observing the
low-luminosity end of the CSO phenomenon.

The speed of $(0.92\pm0.08)~c$ with which the two components separate
is, after IIIZw2, the highest found in a Seyfert galaxy. If the
components are shocks across jets, then the flow speed of the jet
could be even higher. Furthermore, the P.A. in which the separation
increases is in agreement with the relative P.A. of the central and
the NW component, indicating a linear expansion of the source.  The
separation of 282~pc between the components is comparable to the
longest jets found in Seyferts (e.g. \object{NGC~4151},
\citealt{1993MNRAS.263..471P} or \object{Mrk~3},
\citealt{1993MNRAS.264..893K}), and shows that jets in Seyfert
galaxies can propagate large distances while maintaining high speed.

NGC 7674 is remarkable for its broad asymmetric forbidden line
profiles which show highly blue-shifted wings ($2000~{\rm
km\,s^{-1}}$, \citealt{1988MNRAS.234..745U}). These blue wings lie
well outside the galaxy rotation curve and a connection to the AGN or
jet seems likely (\citealt{1988MNRAS.234..745U}), as has been found in
other objects (e.g. \citealt{1988ApJ...326..125W},
\citealt{1998ApJ...496L..75A}). \cite{1992ApJ...387..121W}
investigated the acceleration mechanisms of NLR gas in Seyfert
galaxies. In his sample, Seyferts such as \object{NGC~7674} with
linear radio morphology and radio powers exceeding $10^{22.5}~{\rm
W~Hz^{-1}}$ (\object{NGC~7674} has $2.5\times10^{23}~{\rm W~Hz^{-1}}$,
derived from MERLIN 18~cm integrated flux density) have systematically
broader [O~III] lines, which he attributes to jet material acting on
the NLR gas.

\subsection{\object{NGC~2110}}

\object{NGC~2110} is an elliptical or S0 galaxy. VLA observations at
6~cm and 20~cm with $0.4^{\prime\prime}$ and $1.2^{\prime\prime}$
resolution by
\cite{1983ApJ...264L...7U} show a symmetric, S-shaped radio source
$\sim4^{\prime\prime}$ (600~pc) in extent which was interpreted as a
two-sided, continuous jet interacting with the ISM. This structure was
confirmed in a VLA 3.6~cm image by
\cite{1999ApJ...516...97N} which also resolved finer details in the
jets.\\

\subsubsection{Results}

\begin{figure}[ht!]
\centering
\includegraphics[width=\linewidth]{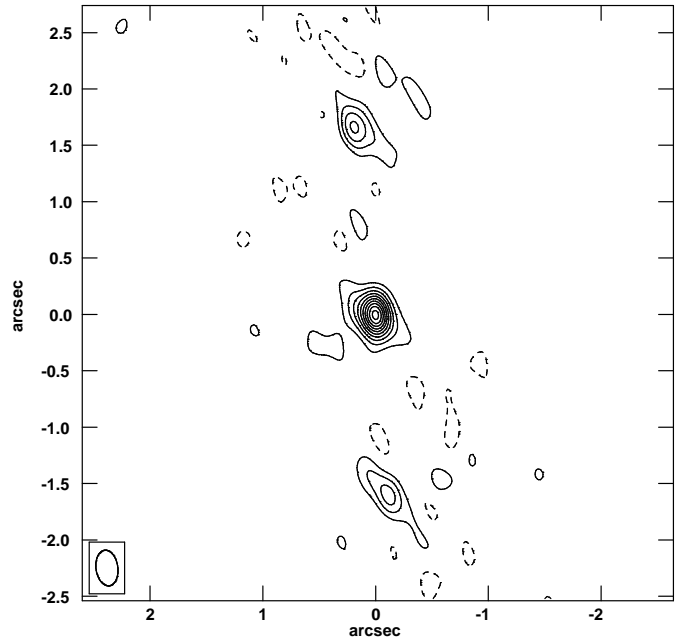}
\caption{\object{NGC~2110} at 18~cm (MERLIN). Offsets are relative to
RA~05:52:11.38254, DEC~$-$07:27:22.5262 (J2000), peak flux density is
31.4~mJy/beam, contours are 1~mJy/beam$\times$($-3$, 3, 6,
9,~$\ldots$,~30), beam size is 318.0~mas$\times$193.4~mas and natural
weighting was used. The $(u,v)$ range was $80-1000~{\rm k}\lambda$.}
\label{fig:ngc2110_merlin_18cm}
\end{figure}

Our MERLIN 18~cm image (Fig.~\ref{fig:ngc2110_merlin_18cm}, tapered to
$0.3\times0.2~{\rm arcsec}$ resolution shows the core and two
components, to north and south, at distances of
$(1.72\pm0.02)^{\prime\prime}$ and $(1.62\pm0.02)^{\prime\prime}$,
respectively, from the central component. These separations are
consistent with those measured by
\cite{1983ApJ...264L...7U} for the brightest parts of the radio
lobes. The extended structure seen in their observations was not
detected by EVN or MERLIN as the surface brightness (360~K) is below
the MERLIN detection threshold (2000~K), except for the bright
hot spots. The MERLIN 6~cm observations yielded only detection of a
point source (not shown).

The total flux density in the MERLIN 6~cm image was 37~mJy, compared
to the single-dish 6~cm flux density of 154~mJy in a $2.4^{\prime}$
beam measured simultaneously at Effelsberg. Thus, three quarters of
the total flux density originates from emission extended over more
than $0.7^{\prime\prime}$ or 75~pc. The total VLA 6~cm flux density
was $(175\pm8)$~mJy in 1983 with a $0.4^{\prime\prime}$ beam
(\citealt{1983ApJ...264L...7U}), which is marginally more than the
single-dish flux density in 2000.

The full-resolution \object{NGC~2110} EVN images at 18~cm and 6~cm
(not shown) display only an unresolved point source at the position of
the VLA observed core. The spectral index of the core was measured
using an EVN 6~cm image tapered to a resolution matching that of the
EVN 18~cm image and the spectrum was found to be rising slightly with
increasing frequency ($\alpha=+0.12$). The MERLIN spectral index of
the core ($+0.20$) was calculated using the full $(u,v)$ range because
tapered 6~cm images had too little $(u,v)$ coverage remaining to yield
reliable images, and no detection was made of the north and south
components.

Since there is no extended emission on the 11~mas to 35~mas scale near
the core, we believe that these spectral indices are unbiased by the
short-baseline-mismatch between the arrays. Observations at higher
resolution carried out by \cite{2000ApJ...529..816M} with the VLBA at
3.6~cm confirm that the core is unresolved down to a resolution of
$2~{\rm mas}$ ($0.3~{\rm pc}$).

\subsubsection{Discussion}

The spectral index of the core in \object{NGC~2110} was estimated by
\cite{1983ApJ...264L...7U} and \cite{1999ApJS..120..209N} from VLA
observations, giving $\alpha^{20}_6=0.0$ and
$\alpha^{20}_{3.6}=-0.47$, respectively. However, both measurements
suffer from the low resolution of the VLA at 20~cm, which does not
allow a clean separation of core and lobes. Therefore, our EVN and
MERLIN observations have allowed an accurate measurement of the core
spectral index for the first time.

\subsection{\object{Mrk~1210}}

Mrk 1210 is an almost face-on, amorphous Seyfert 2 galaxy. It has a
compact NLR (\citealt{1998ApJ...502..199F}), and shows broad emission
lines in polarized light, indicating a hidden type 1 AGN
(\citealt{1992ApJ...397..452T}). \object{Mrk~1210} has also been
classified as a Wolf-Rayet-Galaxy due to its bright He\,II
$\lambda4686$ emission
(\citealt{1998ApJ...501...94S}). \cite{1998ApJ...502..199F} imaged
\object{Mrk~1210} in ${\rm H\alpha}$ and [O\,{\scriptsize III}] with the HST,
to try to resolve an ionization cone, and at 3.6~cm with the VLA, with
resolutions of 100~mas and 200~mas, respectively. In both of those
observations \object{Mrk~1210} appears compact though with a
marginally significant extension towards the southeast in both their
optical and radio observations.

The single-dish 6~cm flux density was 76~mJy in a $4.2^{\prime}$ beam
measured in 1990 (\citealt{1995ApJS...97..347G}) at the Parkes 64~m
telescope.  Mrk 1210 hosts an ${\rm H_2O}$ megamaser with an isotropic
luminosity of $84~L_{\odot}$ (\citealt{1994ApJ...437L..99B}) which
might indicate the presence of an edge-on circumnuclear molecular disk
or torus.\\

\subsubsection{Results}

\begin{figure*}[ht!]
\centering
  \subfigure[MERLIN 6~cm image of \object{Mrk~1210}. Offsets are relative to
  RA~08:04:05.86011, DEC~05:06:49.8992 (J2000), peak flux density is
  24.1~mJy/beam, contours are 1~mJy/beam$\times$($-1$, 1, 3, 5,
  7,~$\ldots$,~23), beam size is 72.7~mas$\times$47.6~mas and uniform
  weighting was used.]{\includegraphics[width=0.4\linewidth]{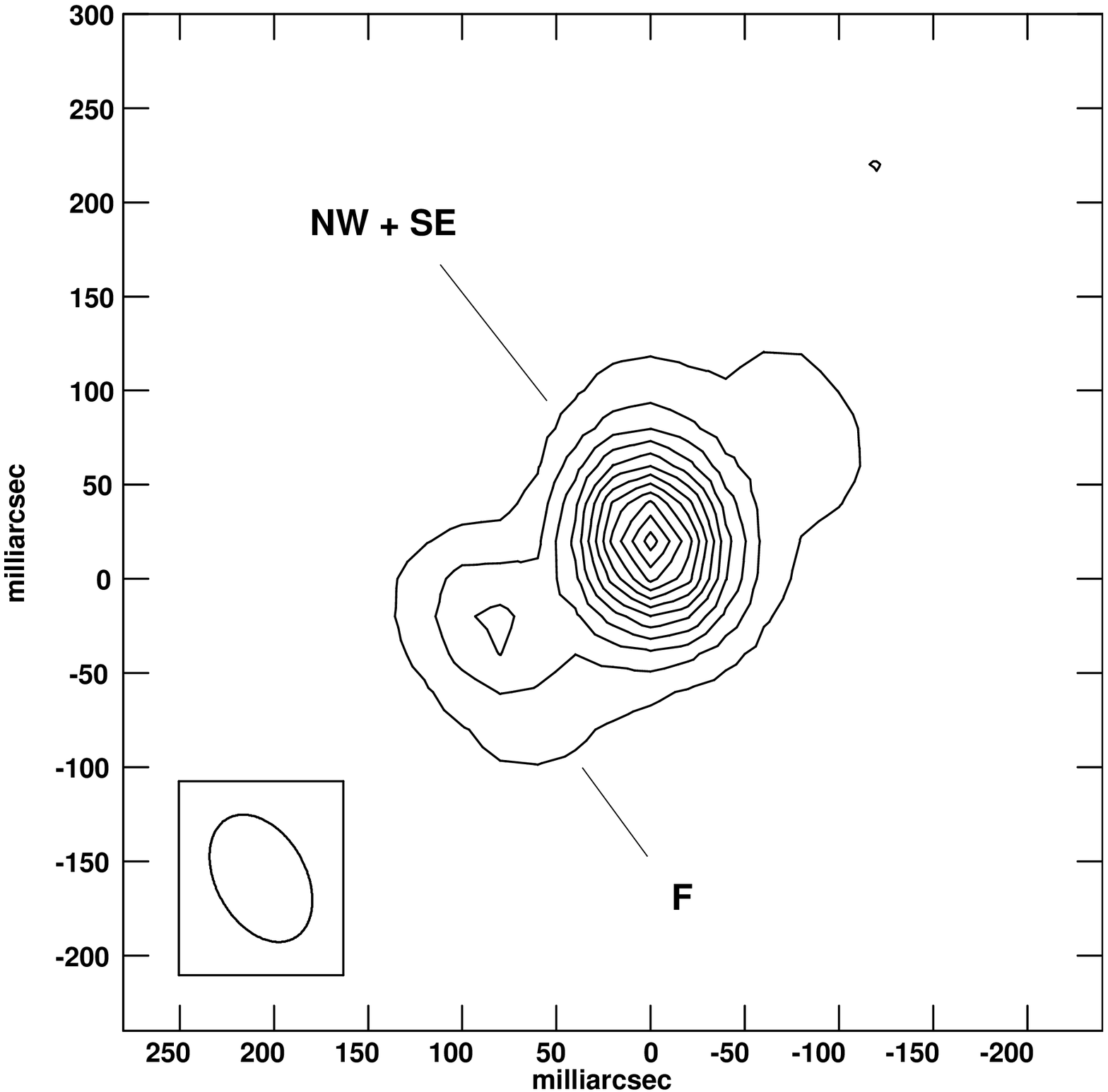}
  \label{fig:mrk1210_merlin_6cm}}
  \subfigure[EVN 18~cm image of \object{Mrk~1210}. Offsets are relative to
  RA~08:04:05.856, DEC~05:06:49.83 (J2000), peak flux density is
  19.8~mJy/beam, contours are 1~mJy/beam$\times$($-1$, 1, 3,
  5,~$\ldots$,~19), beam size is 24.6~mas$\times$11.3~mas and uniform
  weighting was used.]{\includegraphics[width=0.4\textwidth]{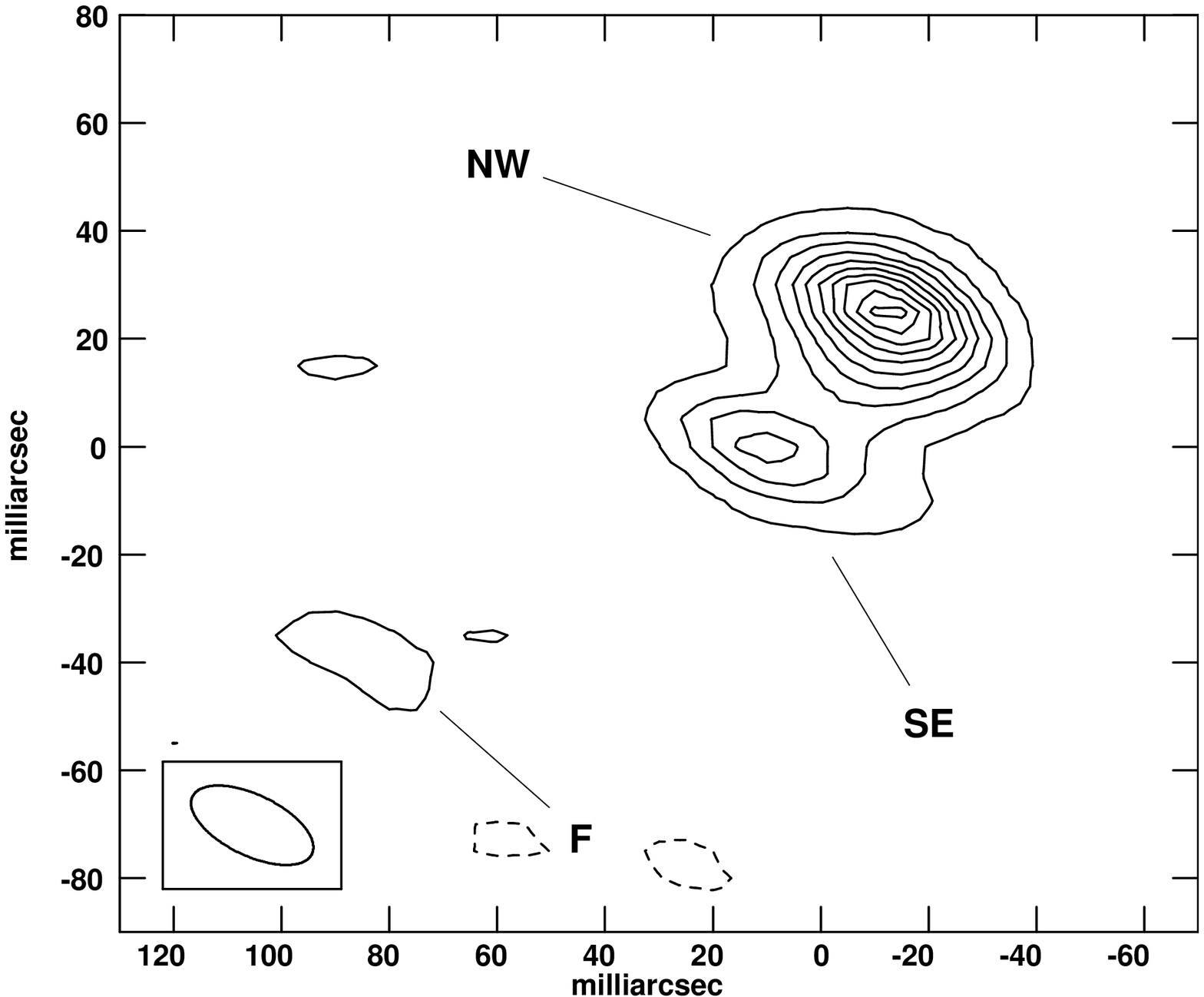}
  \label{fig:mrk1210_evn_18cm}}\\
  \subfigure[VLBA 18~cm image of \object{Mrk~1210}. Offsets are relative to
  RA~08:04:05.841, DEC~05:06:49.72 (J2000), peak flux density is
  21.6~mJy/beam, contours are 1~mJy/beam$\times$($-1$, 1, 3,
  5,~$\ldots$,~21), beam size is 18.6~mas$\times$14.2~mas and natural
  weighting was used.]{\includegraphics[width=0.4\linewidth]{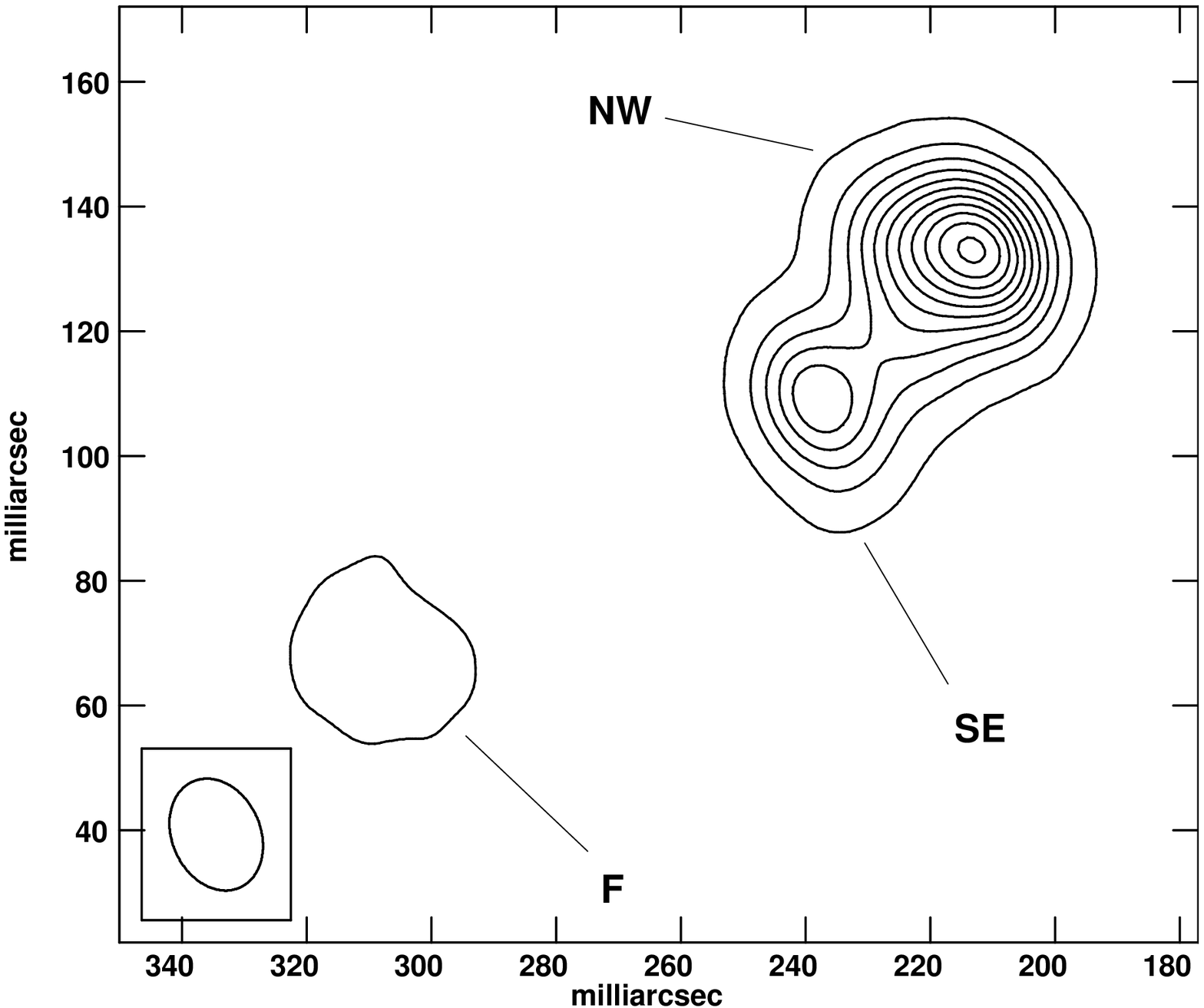}
  \label{fig:mrk1210_vlba_18cm}}
  \subfigure[VLBA 6~cm image of \object{Mrk~1210}. Offsets are relative to
  RA~08:04:05.841, DEC~05:06:49.72 (J2000), peak flux density is
  2.5~mJy/beam, contours are 0.1~mJy/beam$\times$($-3$, 3,
  6,~$\ldots$,~24), beam size is 3.2~mas$\times$1.5~mas and uniform
  weighting was used.]{\includegraphics[width=0.4\textwidth]{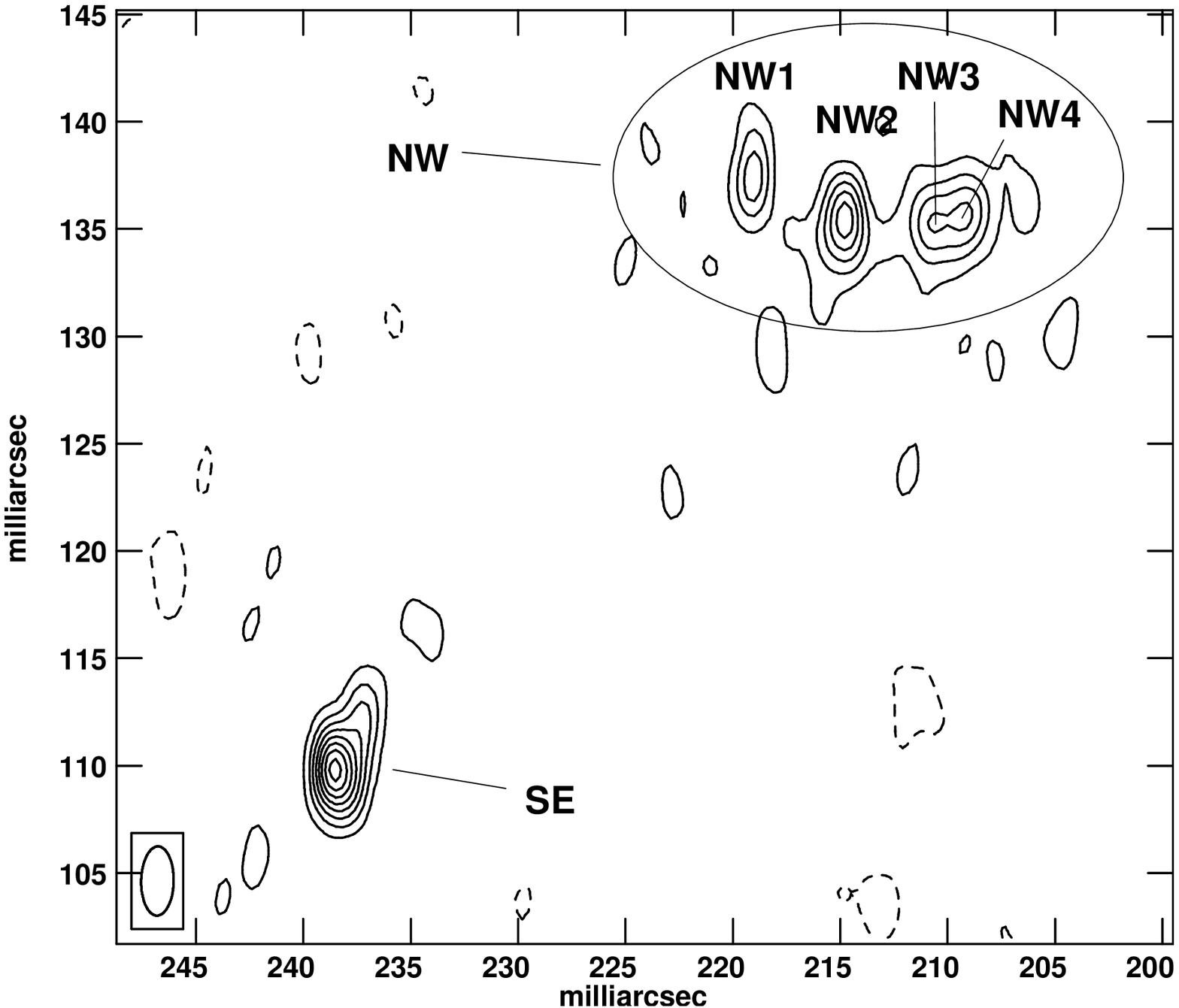}
  \label{fig:mrk1210_vlba_6cm}}\\
  \caption{Mrk 1210 images.}
\end{figure*}

The MERLIN 18~cm image (not shown) displays an unresolved point source
with a possible weak halo to the north-west.  The MERLIN 6~cm image
(Fig.~\ref{fig:mrk1210_merlin_6cm}) reveals a bright, unresolved
central component with a weak extension to the north-west, and a third
component to the far south-east, which we name F. The total MERLIN
6~cm flux density was 39~mJy compared to the 6~cm single-dish flux
density of $54~{\rm mJy}$ in a $2.4^{\prime}$ beam measured
simultaneously at Effelsberg on the same day, and so the MERLIN image
has recovered most of the flux density.

The EVN and VLBA 18~cm images in Fig.~\ref{fig:mrk1210_evn_18cm} and
\ref{fig:mrk1210_vlba_18cm} reveal a bright compact object (named NW)
and a weaker component towards the south-east (named SE) at a distance
of $(32.7\pm1.0)~{\rm mas}$ (8.6~pc) from the NW object. We marginally
detect a third component at a distance of $\approx117~{\rm mas}$
(30.6~pc) from NW which is spatially coincident with the weak object F
in the MERLIN 6~cm image. This component is not formally significant
but is present in both images and so is probably real and could be the
continuation of the radio ejecta. The EVN 6~cm data (not shown) yield
only a marginal detection of the strongest point source. The 6~cm VLBA
image (Fig.~\ref{fig:mrk1210_vlba_6cm}) resolves the brightest source
into an arc of four components and also resolves the bright component,
named SE. The total flux density recovered by the full-resolution EVN
and VLBA at 18~cm was 50~mJy and 49~mJy, and at 6~cm was 10~mJy and
9.3~mJy, respectively. The total 6~cm flux density measured by the
VLBA in a beam tapered to the 18~cm resolution is 14~mJy, which is
still only $35~\%$ of the total MERLIN flux density. So most of the
galaxy's 6~cm emission comes from scales between 3~pc (the largest
angular scale to which the EVN is sensitive) and MERLIN's beam size of
16~pc, and only little from even larger scales.

Both the NW and SE components have steep spectra, with $\alpha=-1.26$
and $\alpha=-0.78$ in our VLBA images. In our EVN images, the NW
component has $\alpha=-1.19$ provided that the weak detection in our
EVN 6~cm observations is the NW component. Since all components have
steep spectral indices, the core identification is best done based on
proper motions. Our observations here provide the first epoch absolute
positions and a second epoch should follow after at least five years
to allow a reliable detection of proper motion.

The position angle of the SE component with respect to the NW
component ($122\pm5^{\circ}$) is perpendicular to the optical
polarization vector (P.A. $29^{\circ}$) measured by
\cite{1995ApJ...440..578T}, as is generally the case in Seyfert 2s.

\subsubsection{Discussion}

We suggest three scenarios for the radio structure in \object{Mrk~1210}.

{\bf A free-free emitting disc?}~~The component configuration reminds
us of \object{NGC~1068} in which the core component S1 at the jet base is
resolved into a 15~mas (1~pc)-long structure perpendicular to the jet
axis in the sensitive 3.6~cm VLBA observation by
\cite{1997Natur.388..852G}. Surrounding this component is an edge-on
disc of ${\rm H_2O}$ masers whose rotation curve implies a compact
central mass of $1.5\times10^7~M_{\odot}$
(\citealt{1997Ap&SS.248..261G}). In that case, the 3.6~cm continuum
emission was identified as free-free emission from the ionized inner
edge of the torus, based on its flat spectrum, relatively large
extent, orientation perpendicular to the jet, and low brightness
temperature ($10^6~{\rm K}$), the last ruling out synchrotron
self-absorption as the cause of the flat spectrum. In
\object{Mrk~1210}, the bright arc of four components suggests an
edge-on disk, whose existence is further supported by the presence of
nuclear water maser emission (although the precise location of the
masers is not known). The integrated radio spectrum of component NW is
$\alpha^{18}_{6}=-1.2$ using the EVN and $\alpha^{18}_{6}=-1.3$ using
the VLBA. This favours optically thin synchrotron radiation rather
than flat-spectrum free-free emission.  The brightness temperature
measured with the VLBA at 18~cm is $9.2\times10^7~{\rm K}$ and with
the EVN at 18~cm is $7.1\times10^7~{\rm K}$, significantly higher than
in NGC~1068. Thus, although the shape of the radio emitting regions in
component NW is suggestive, the structure cannot be free-free emission
from the inner edge of a torus like in NGC~1068.

{\bf Synchrotron emission from a torus or extended accretion
disk?}~~The arc of four components in the NW component could be
synchrotron emission from an edge-on viewed torus or accretion
disc. However, accretion disks are normally thought to be much smaller
and neither accretion disks nor tori are known as synchrotron
emitters.

{\bf A core and outflow?}~~Perhaps component SE hosts the nucleus.
The resolved arc of four components in NW then could be bright knots
due to a shock in an extended, low brightness temperature radio
outflow, with similar conditions to those in \object{NGC~5506},
NGC~1068 or Mrk~231. But both the SE and the F component lack a flat
spectrum that would indicate the nucleus.

Further clues might come from absolute proper motion measurements in
the future, or from VLBI imaging of the water maser emission.

\section{Statistics}

The range of radio properties displayed by Seyfert galaxies has been
explored extensively based on WSRT, VLA or ATCA observations at
20~cm, 6~cm or 3.6~cm at arcsec resolution by
\cite{1978AaA....64..433D}, \cite{1984ApJ...285..439U,
1989ApJ...343..659U} \cite{1987MNRAS.228..521U},
\cite{1990ApJS...72..551G}, \cite{1996ApJ...473..130R},
\cite{1999ApJ...516...97N}, \cite{1999AAS..137..457M}, 
\cite{2001ApJ...558..561U}, and \cite{VirLal2001}.  They studied the
range of radio luminosities, radio source sizes, morphologies,
alignment between jet and host galaxy, and spectral indices, and
looked for dependence of properties on Seyfert type. With the
exception of \cite{2001ApJ...558..561U}, who found 50\,\% of
low-luminosity Seyferts to have a flat spectrum core, they all found
that most Seyfert galaxies have steep, optically-thin synchrotron
radio spectra, with 5\,\% to 35\,\% (depending on the sample) showing
a flat-spectrum core, much less than the rate of occurrence of
flat-spectrum cores in powerful radio galaxies.  No difference was
found between the spectral index distributions of the two Seyfert
types.

However, the $0.3^{\prime\prime}$ to several arcsecs beams used in
those studies do not allow details of nuclear structure to be
separated or the distribution of jet bending angles to be
constructed. Further, they would probably blend together multiple
radio components within the nucleus and could hide the presence of
more absorbed cores.  Our VLBI observations provide enough resolution
to measure separately the emission from individual sub-pc-scale
components, and so can find parsec-scale absorbers more
efficiently. However, our sample size is too small for any general
conclusions.

Many further VLBI observations of Seyfert galaxies are available in
the literature but no statistical summary from those observations has
been presented.  Thus, we made a literature search for VLBI
observations of Seyferts to supplement our sample. The resulting list
in Table~\ref{tab:seyferts} contains, we believe, all VLBI imaging
observations of AGN classified as Seyferts by \cite{Veron2001},
published as of August 2003 with a linear resolution of $<10~{\rm
pc}$, although it is not impossible that we have inadvertently omitted
some works.  This sample consists mostly of the nearest and best-known
Seyferts.  They were often selected for observing because they show
bright radio nuclei at arcsecond resolution, and so the sample is not
complete in any sense.  Large, uniform samples of Seyfert galaxies are
being observed with the VLBA by \cite{Murray1999} and
\cite{2002AAS...200.4507S}, but the results are not yet
available. Proper motions in Seyferts are being presented by Roy et
al. (in prep.) and we confine ourselves here to spectral indices and
jet bending. 

\begin{figure}[ht!]
\centering
  \subfigure[]{
  \includegraphics[width=0.4\textwidth, angle=270]{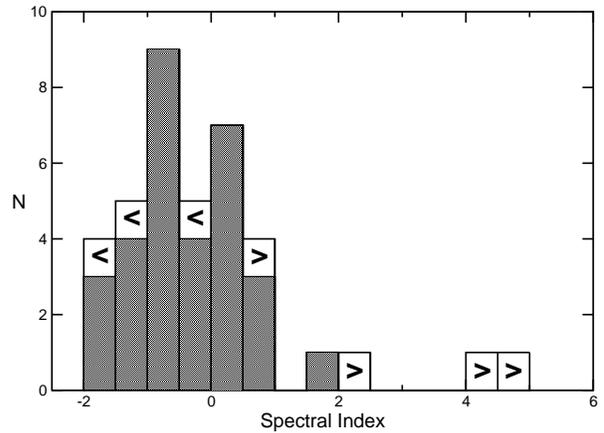}
  \label{fig:histogram_SI}}
  \subfigure[]{
  \includegraphics[width=0.4\textwidth, angle=270]{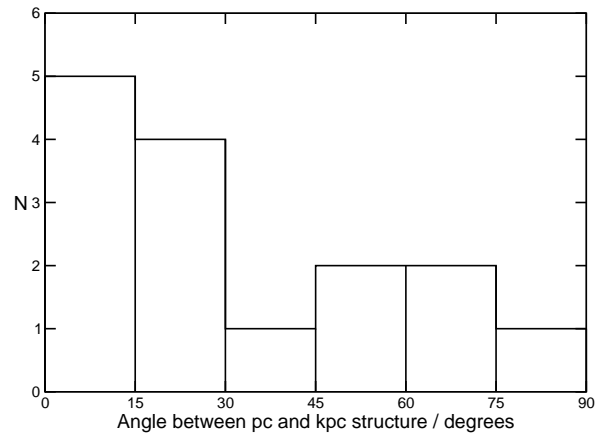}
  \label{fig:histogram_PA}}\\
  \caption{{\it Upper panel}: Histogram of spectral indices between 18 cm
  and 6 cm for the Seyfert sample in Table~\ref{tab:spectra}.  Upper
  and lower limits are indicated by arrows. {\it Lower
  panel:} Distribution of angles between pc- and kpc-scale
  structure. For comparison with data from radio galaxies, the data
  from Table~\ref{tab:spectra-stats2} have been reflected onto the range
  $0^{\circ}$ to $90^{\circ}$.}
\end{figure}

\subsection{Fraction of Flat-Spectrum Components}

We selected from Table~\ref{tab:seyferts} a subsample consisting of
the Seyfert galaxies for which dual-frequency VLBI observations have
been made, including the four objects presented in this paper, to
study the component spectral properties at high resolution. This
subsample is listed in Table~\ref{tab:spectra}.

Of the 16 objects in the subsample, a majority of 13 (81~\%) has at
least one flat or inverted spectrum VLBI component with $\alpha^{18}_6
> -0.3$.  This fraction is 60\,\% larger than that measured at lower
resolution by \cite{2001ApJS..133...77H} for the Palomar Seyferts and
four or more times larger than the fraction measured at arcsec
resolution for the other samples (Table~\ref{tab:spectra-stats}).  The
difference is highly significant for most samples, having a
probability of $<0.1~\%$ of occurring at random.  The larger fraction
of flat- and inverted-spectrum components found with VLBI is not
unexpected since increasing resolution reduces the component blending.

\subsection{Spectral Index Distributions}

The distribution of $\alpha^{18}_6$ from Table~\ref{tab:spectra} is
shown in Fig.~\ref{fig:histogram_SI}.  The spectral indices lie
between $-1.8$ and $+4.6$ with a median of $-0.40$ (using the
Kaplan-Meier estimator constructed by ASURV rev 1.2,
\citealt{1992adass...1..245L} to account for lower limits and
discarding upper limits).  The distribution shows a tail towards
positive spectral index due to four components in NGC~1068,
\object{NGC~2639} and \object{NGC~3079} that have strong low-frequency
absorption.  In the case of two components in NGC~3079, the spectral
indices are greater than the maximum $\alpha^{18}_6 = +2.5$ that can
be produced by synchrotron self absorption and so is most likely due
to free-free absorption, probably by dense circumnuclear plasma
ionized by the AGN.

The tail of extreme positive spectral indices that we see in our
sample is not seen at arcsec resolution in the works mentioned above,
probably due to component blending.  We made a statistical comparison
of our VLBI spectral indices to arcsec resolution spectral indices
using the Kolmogorov-Smirnov test and chi-squared homogeneity test by
treating the lower and upper limits as detections, and by applying
survival analysis two-sample tests.  The results are summarized in
Table~\ref{tab:spectra-stats2}.  Our spectral index distribution was
significantly different (i.e. level of significance was less than the
critical value of 5\,\%) from all the comparison samples according to
all tests performed, with the exception of the Palomar Seyferts by
\cite{2001ApJS..133...77H}, which was found to be significantly
different in only the chi-squared test. This excess of flat-spectrum
sources in the Palomar Seyfert sample was already noticed by
\cite{2001ApJS..133...77H}.

Separating the VLBI measurements by Seyfert type, we found the median
$\alpha^{18}_6$ for the Sy 1s was $-0.75$ and for the Sy 2s was
$-0.23$, which were not significantly different under the survival
analysis two-sample tests ($P=6\,\%$ to $12\,\%$ that the
spectral-index distributions came from the same parent distribution)
or under a Kolmogorov-Smirnov test ($P=6.1\,\%$ to come from the same
parent distribution), treating only lower limits as detections.

Despite the statistical insignificance of a difference between Seyfert
types, we notice that the most inverted spectra all occur in objects
that have evidence for an absorber. NGC~3079 is a type 2 Seyfert,
NGC~1068 and Mrk~348 are type 2 Seyferts with type 1 spectra in
polarized light and NGC~2639 shows strong X-ray absorption
(\citealt{1998ApJ...505..587W}). The inner edge of such an absorber is
expected to be ionized by UV emission from the AGN which would produce
free-free absorption and hence produce inverted spectra preferentially
in the type 2s.  All four objects with inverted radio spectra also
host ${\rm H_2O}$ masers, which is also consistent with an orientation
effect if, for example, the radio core is viewed through a molecular
disk or torus, which produces ${\rm H_2O}$ maser emission.

\subsection{Jet Misalignment}
\label{sec:misalignment}

Kiloparsec-scale linear radio structures in Seyfert galaxies are
thought to be energized by jets or outflows originating at the AGN
core, and so one might expect pc and kpc-scale radio structures to be
aligned.  Misalignments could be caused by changes in the ejection
axis or by pressure gradients in the ISM.  To look for such effects,
we compiled kpc-scale radio observations of the VLBI Seyfert sample,
where available. The resulting sample of 21 objects in
Table~\ref{tab:misalignment} is, we believe, a complete list of
Seyferts with both pc-scale and kpc-scale observations published,
although seven of these objects either have no clear extended
structure on one of the scales or the structure was deemed to be
affected by processes on scales $>1\,{\rm kpc}$, e.g., buoyancy
forces. In one object, \object{NGC~1167}, the double-sided kpc-scale
emission is bent, so we quote both.  This sample has a large overlap
with the sample in Table~\ref{tab:spectra} and, likewise, is not
complete in any astronomical sense.  Biases such as selection for
bright radio nuclei are probably present, but whether they affect the
distribution of jet bending angle is unclear.  Doppler boosting would
favour jets that point towards us, but the jet speeds so far measured
in Seyferts are low and boosting effects should be minor.

The distribution of misalignment angles from
Table~\ref{tab:misalignment} is shown in Fig.~\ref{fig:histogram_PA}.
Misalignment in Seyfert jets between pc and kpc-scales is common, with
5 out of 15 objects bending by $45^{\circ}$ or more.  The null
hypothesis that the jet bending-angle distribution was drawn from a
parent population of aligned jets was rejected by a chi-squared test
at the 2\,\% level of significance.  The null hypothesis that the
distribution was drawn from a uniformly distributed parent population
could not be rejected by a chi-squared test, which returned a 25\,\%
level of significance.

We compared the bending angles in Seyferts to those in radio-loud
objects using a comparison sample of core-dominated radio-loud objects
compiled from Table~6 in \cite{1988ApJ...328..114P}, and Table~1 in
\cite{1993ApJ...411...89C}, which have both VLBI and VLA or
MERLIN observations at 5~GHz.  Those objects show a surprising bimodal
distribution of jet bending angle, in which the pc-scale jets tend to
align with or to be perpendicular to the kpc-scale structure.  We
compared our jet bending angle distribution to the distribution in
radio-loud objects using a Kolmogorov-Smirnov test and found no
significant difference, the probability of being drawn from the same
parent distribution being 14\,\%.  However, the Seyfert sample is
small and the distribution for the Seyferts was shown in the previous
paragraph to be consistent with a uniform distribution.

The frequent misalignments seen in Seyfert galaxy jets could be
evidence either for changes in the jet ejection angle and hence of the
accretion disk, perhaps by a mechanism such as radiation-driven
warping (\citealt{1997MNRAS.292..136P}), or for bending due to
pressure gradients in the ISM (collision or buoyancy) through which
the jets propagate.

We compared the extended pc-scale radio structures to the host galaxy
rotation axes, assuming that the host galaxy is circular and the
rotation axis vector is projected onto the galaxy's minor axis, which
can be measured from optical images. For all spiral galaxies with
pc-scale radio structure, we used the major axes from
\cite{1991trcb.book.....D}, subtracted $90^{\circ}$ and reflected the
angles onto the range $0^{\circ}$ to $90^{\circ}$. Four objects were
not listed in \cite{1991trcb.book.....D}, so we used data from
\cite{Schmitt2000} in two cases and we measured the major axes from
Hubble Space Telescope archival data in the other two cases. We
performed a Spearman rank correlation test and found a correlation
coefficient of 0.07, corresponding to a 82\,\% significance and so
there is no correlation between the directions of the pc-scale
structures and the host galaxy rotation axes. However, the sample was
small (13 objects) and therefore the test is not strong.

\section{Conclusions}

We have observed the Seyfert galaxies \object{NGC~5506},
\object{NGC~7674}, \object{NGC~2110} and \object{Mrk~1210} at 18~cm
and 6~cm with the EVN and MERLIN.  \object{NGC~5506} and
\object{Mrk~1210} were additionally observed with the VLBA.  These
observations provided high angular resolution and good surface
brightness sensitivity, spectral index measurements, and up to four
epochs of observations for proper motion measurements.  Our
conclusions are as follows.\\

1. The structure seen on the tens-of-parsec-scale in \object{NGC~7674}
shows two bright, steep-spectrum components that might be the working
surfaces where double-sided jets ejected from an (unseen) AGN core
located between them impact the interstellar medium.
Low-surface-brightness emission was seen extending from the hot spots
in the direction of VLA-observed emission by \cite{Momjian2003}.  The
diffuse emission could be from material that has flowed through the
hotpots and was then either deflected or rose buoyantly in the
pressure gradient in the ISM.  Alternatively, the bend after the hot
spot might be caused by accretion disk precession, which might move
the hot spots sideways along a wall of ISM leaving behind a trail of
fading extended plasma.  These structures are reminiscent of those
seen in compact symmetric objects, so we suggest that
\object{NGC~7674} is a low-luminosity member of the CSO family.

\object{NGC~5506} shows two bright radio components separated by 3.5\,pc with
low-surface-brightness emission on larger scales extending towards the
VLA-scale emission. The brightest compact component (B0) has a flat
radio spectrum ($\alpha=0.06$) and could represent the nucleus.

In \object{NGC~2110} and \object{Mrk~1210}, the overall structure is a
triple radio source. For \object{NGC~2110}, the flat spectrum
($\alpha=0.12$) and the symmetric structure of the VLA-observed jets
and the components at the ends of these jets (also seen in our MERLIN
image) show conclusively that the central source represents the
nucleus. In \object{Mrk~1210}, both the SE and the NW components have
steep radio spectra and here the location of the nucleus is unclear.

If jets in Seyferts are initially relativistic and are then disrupted
by interaction with the ISM, we might expect significant faint,
diffuse radio emission. Such faint emission (brighter than $2~{\rm
mJy\,beam^{-1}}$) was seen in only \object{NGC~5506}.\\

2. A proper motion of $(0.92\pm0.07)~c$ was measured in
\object{NGC~7674}, and a $3~\sigma$ upper limit of $0.50~c$ was found
for the motion of B1 in \object{NGC~5506} with respect to B0.  From
the high speed measured in NGC 7674 between components 282~pc apart we
derived a dynamical age of 1000~yr.  This speed and age is similar to
those seen in CSOs.  That the components have high speed after
propagating hundreds of parsecs shows that the ISM in this object has
not stopped the jet very effectively.  The speed measured in
\object{NGC~7674} is faster than those measured in other Seyferts by
\cite{1998ApJ...496..196U}, \cite{1999ApJ...517L..81U} and
\cite{2000evn..proc....7R}, except for IIIZw2, though is slow compared
to the speeds at equivalent scales in powerful radio galaxies.  This
suggests, but does not prove, that the jets are launched with an
intrinsically lower speed than in radio galaxies.  Further support
comes from the preference of Seyfert nuclei for early-type spirals,
which have relatively low-density ISM even close to the nucleus, and
hence it seems unlikely that the jets are significantly slowed.\\

3. We mentioned in the introduction the many possible reasons for low
jet power in Seyferts that have been suggested in the literature.  Our
observations of proper motions favour an intrinsic over an extrinsic
cause since the ISM has not stopped the component in
\object{NGC~7674}, and no emission from disrupted jets was seen in
three of four Seyferts. Any extrinsic deceleration must happen on
scales less than 3.5~pc.  (However, the bends in the radio sources at
distances of 3.5 pc and 282~pc in \object{NGC~5506} and
\object{NGC~7674} show that the putative jets may nevertheless feel an
influence from the NLR.)  Of the intrinsic causes mentioned in the
introduction, we cannot constrain the role of black-hole spin or
magnetic field strength in the accretion disc.  The expanding plasmon
model of \cite{1985MNRAS.214..463P} predicts that the radio components
expand into an ISM of density $1~{\rm cm^{-3}}$ at speeds initially of
$\ge 10^4~{\rm km\,s^{-1}}$ and would reach a size of 30~pc after
2000~yr by which time the expansion speed would have been reduced to
$5000~{\rm km\,s^{-1}}$.  At this speed and at the distance of
\object{NGC~5506} over the 6.25~yr span of our observations, the
component diameters would increase by 0.3~mas, or, in
\object{NGC~7674}, over 14.58~yr by 0.1 mas. An increase by this
amount is, unfortunately, much smaller than could be measured with our
data.\\

4. Spectral indices: We asked in the introduction whether Seyfert
nuclei show a single compact synchrotron self-absorbed core component
at base of a collimated jet, like scaled-down radio galaxies.  We
found flat- or absorbed-spectrum components in \object{NGC~5506} and
\object{NGC~2110} with brightness temperatures of $\ge
3\times10^7~{\rm K}$.  The components in \object{NGC~5506} have no
clear axis of symmetry and the flat spectrum could be caused by either
free-free absorption or synchrotron self-absorption.  The morphology
of \object{NGC~2110} shows collimated jets on VLA scales and so is the
only object in our sample with a flat-spectrum VLBI core and
large-scale jets.  Thus, the classical conically-expanding
Blandford-K\"onigl jet structure (\citealt{Blandford1979}, see also
\citealt{Falcke1995}) is sometimes but not often seen in our Seyfert
galaxies.\\

5. Statistics of selected radio properties: To improve the reliability
of any conclusions we enlarged our sample by drawing Seyferts from the
literature that were observed with VLBI and found that flat- or
absorbed- spectrum cores were present in 13 of 16 cases.  The most
strongly absorbed components tend to be in objects that also show
other evidence for absorption, such as water maser emission or large
X-ray absorption columns.  In those cases, according to the Seyfert
unification schemes, the flat-spectrum components would be viewed
through an edge-on obscuring disk or torus, the ionized inner edge of
which could be causing free-free absorption.  If so, the situation in
those objects is different from the Blandford-K\"onigl jet model.
However the case for free-free absorption is unambiguous in only one
galaxy (NGC~3079) for which the spectral index is $>+2.5$.  The other
12 Seyferts have $-0.3<\alpha<+2.5$ and so spectral turnovers could be
produced by either synchrotron self-absorption or free-free
absorption.

Where extended structures are present in the enlarged sample (14
Seyferts), the position angle differences between pc-scale and
kpc-scale radio emission were found to be uniformly distributed
between $0^{\circ}$ and $90^{\circ}$.  Such bends could be due to
changes in the jet ejection axis or due to pressure gradients in the
ISM. No correlation was found between the axis of pc-scale radio
structure and the rotation axis of the host galaxy.

\begin{acknowledgements}
We thank the referee (R. Morganti) for the time and care that she
invested in critiquing our manuscript.  Her numerous suggestions have
improved the paper enormously. We wish to thank R.~W.~Porcas for
providing 15 year old $T_{\rm sys}$ measurements for the amplitude
calibration of the 1985 EVN observations and Jim Ulvestad for help
with the reduction of the \object{Mrk~1210} VLBA data. The Effelsberg
telescope is operated by the Max-Planck-Institut f\"ur
Radioastronomie. The European VLBI Network is a joint facility of
European, Chinese and other radio astronomy institutes funded by their
national research councils. The VLBA is an instrument of the National
Radio Astronomy Observatory, a facility of the National Science
Foundation, operated under cooperative agreement by Associated
Universities, Inc. This research was supported in part by NASA through
grant NAG81755 to the University of Maryland and has made use of the
NASA/IPAC Extragalactic Database (NED) which is operated by the Jet
Propulsion Laboratory, California Institute of Technology, under
contract with the National Aeronautics and Space Administration.
\end{acknowledgements}

\bibliography{h4137}

\begin{thebibliography}{117}
\expandafter\ifx\csname natexlab\endcsname\relax\def\natexlab#1{#1}\fi

\bibitem[{{Alef} {et~al.}(1996){Alef}, {Wu}, {Preuss}, {Kellermann}, \&
  {Qiu}}]{1996AaA...308..376A}
{Alef}, W., {Wu}, S.~Y., {Preuss}, E., {Kellermann}, K.~I., \& {Qiu}, Y.~H.
  1996, \aap, 308, 376

\bibitem[{{Antonucci} \& {Miller}(1985)}]{Antonucci1985}
{Antonucci}, R.~R.~J. \& {Miller}, J.~S. 1985, \apj, 297, 621

\bibitem[{{Axon} {et~al.}(1998){Axon}, {Marconi}, {Capetti}, {Maccetto},
  {Schreier}, \& {Robinson}}]{1998ApJ...496L..75A}
{Axon}, D.~J., {Marconi}, A., {Capetti}, A., {et~al.} 1998, \apjl, 496, L75

\bibitem[{{Barvainis} \& {Lonsdale}(1998)}]{1998AJ....115..885B}
{Barvainis}, R. \& {Lonsdale}, C. 1998, \aj, 115, 885

\bibitem[{{Bicknell} {et~al.}(1998){Bicknell}, {Dopita}, {Tsvetanov}, \&
  {Sutherland}}]{1998ApJ...495..680B}
{Bicknell}, G.~V., {Dopita}, M.~A., {Tsvetanov}, Z.~I., \& {Sutherland}, R.~S.
  1998, \apj, 495, 680

\bibitem[{{Blandford} \& {Konigl}(1979)}]{Blandford1979}
{Blandford}, R.~D. \& {Konigl}, A. 1979, \apj, 232, 34

\bibitem[{{Bower} {et~al.}(1995){Bower}, {Wilson}, {Morse}, {Gelderman},
  {Whittle}, \& {Mulchaey}}]{1995ApJ...454..106B}
{Bower}, G., {Wilson}, A., {Morse}, J.~A., {et~al.} 1995, \apj, 454, 106

\bibitem[{{Braatz} {et~al.}(1994){Braatz}, {Wilson}, \&
  {Henkel}}]{1994ApJ...437L..99B}
{Braatz}, J.~A., {Wilson}, A.~S., \& {Henkel}, C. 1994, \apjl, 437, L99

\bibitem[{{Browne} {et~al.}(1998){Browne}, {Wilkinson}, {Patnaik}, \&
  {Wrobel}}]{1998MNRAS.293..257B}
{Browne}, I.~W.~A., {Wilkinson}, P.~N., {Patnaik}, A.~R., \& {Wrobel}, J.~M.
  1998, \mnras, 293, 257

\bibitem[{{Brunthaler} {et~al.}(2000){Brunthaler}, {Falcke}, {Bower}, {Aller},
  {Aller}, {Ter{\" a}sranta}, {Lobanov}, {Krichbaum}, \&
  {Patnaik}}]{2000AaA...357L..45B}
{Brunthaler}, A., {Falcke}, H., {Bower}, G.~C., {et~al.} 2000, \aap, 357, L45

\bibitem[{{Capetti} {et~al.}(1997){Capetti}, {Axon}, \&
  {Macchetto}}]{1997ApJ...487..560C}
{Capetti}, A., {Axon}, D.~J., \& {Macchetto}, F.~D. 1997, \apj, 487, 560

\bibitem[{{Carilli} \& {Taylor}(2000)}]{2000ApJ...532L..95C}
{Carilli}, C.~L. \& {Taylor}, G.~B. 2000, \apjl, 532, L95

\bibitem[{{Cecil} {et~al.}(2000){Cecil}, {Greenhill}, {DePree}, {Nagar},
  {Wilson}, {Dopita}, {P{\' e}rez-Fournon}, {Argon}, \&
  {Moran}}]{2000ApJ...536..675C}
{Cecil}, G., {Greenhill}, L.~J., {DePree}, C.~G., {et~al.} 2000, \apj, 536, 675

\bibitem[{{Colbert} {et~al.}(1996){Colbert}, {Baum}, {Gallimore}, {O'Dea}, \&
  {Christensen}}]{1996ApJ...467..551C}
{Colbert}, E.~J.~M., {Baum}, S.~A., {Gallimore}, J.~F., {O'Dea}, C.~P., \&
  {Christensen}, J.~A. 1996, \apj, 467, 551

\bibitem[{{Conway} \& {Murphy}(1993)}]{1993ApJ...411...89C}
{Conway}, J.~E. \& {Murphy}, D.~W. 1993, \apj, 411, 89

\bibitem[{{de Bruyn} \& {Wilson}(1978)}]{1978AaA....64..433D}
{de Bruyn}, A.~G. \& {Wilson}, A.~S. 1978, \aap, 64, 433

\bibitem[{{de Vaucouleurs} {et~al.}(1991){de Vaucouleurs}, {de Vaucouleurs},
  {Corwin}, {Buta}, {Paturel}, \& {Fouque}}]{1991trcb.book.....D}
{de Vaucouleurs}, G., {de Vaucouleurs}, A., {Corwin}, H.~G., {et~al.} 1991,
  {Third Reference Catalogue of Bright Galaxies} (Volume 1-3, XII, 2069 pp.~7
  figs..~ Springer-Verlag Berlin Heidelberg New York)

\bibitem[{{Duric} {et~al.}(1983){Duric}, {Seaquist}, {Crane}, {Bignell}, \&
  {Davis}}]{1983ApJ...273L..11D}
{Duric}, N., {Seaquist}, E.~R., {Crane}, P.~C., {Bignell}, R.~C., \& {Davis},
  L.~E. 1983, \apjl, 273, L11

\bibitem[{{Falcke} \& {Biermann}(1995)}]{Falcke1995}
{Falcke}, H. \& {Biermann}, P.~L. 1995, \aap, 293, 665

\bibitem[{{Falcke} {et~al.}(2000){Falcke}, {Nagar}, {Wilson}, \&
  {Ulvestad}}]{2000ApJ...542..197F}
{Falcke}, H., {Nagar}, N.~M., {Wilson}, A.~S., \& {Ulvestad}, J.~S. 2000, \apj,
  542, 197

\bibitem[{{Falcke} {et~al.}(1998){Falcke}, {Wilson}, \&
  {Simpson}}]{1998ApJ...502..199F}
{Falcke}, H., {Wilson}, A.~S., \& {Simpson}, C. 1998, \apj, 502, 199

\bibitem[{{Fomalont} {et~al.}(2000){Fomalont}, {Frey}, {Paragi}, {Gurvits},
  {Scott}, {Taylor}, {Edwards}, \& {Hirabayashi}}]{2000ApJS..131...95F}
{Fomalont}, E.~B., {Frey}, S., {Paragi}, Z., {et~al.} 2000, \apjs, 131, 95

\bibitem[{{Gallimore} {et~al.}(1997){Gallimore}, {Baum}, \&
  {O'Dea}}]{1997Natur.388..852G}
{Gallimore}, J.~F., {Baum}, S.~A., \& {O'Dea}, C.~P. 1997, \nat, 388, 852

\bibitem[{{Giovannini} {et~al.}(2001){Giovannini}, {Cotton}, {Feretti}, {Lara},
  \& {Venturi}}]{2001ApJ...552..508G}
{Giovannini}, G., {Cotton}, W.~D., {Feretti}, L., {Lara}, L., \& {Venturi}, T.
  2001, \apj, 552, 508

\bibitem[{{Giovannini} {et~al.}(1990){Giovannini}, {Feretti}, \&
  {Comoretto}}]{1990ApJ...358..159G}
{Giovannini}, G., {Feretti}, L., \& {Comoretto}, G. 1990, \apj, 358, 159

\bibitem[{{Giuricin} {et~al.}(1990){Giuricin}, {Mardirossian}, {Mezzetti}, \&
  {Bertotti}}]{1990ApJS...72..551G}
{Giuricin}, G., {Mardirossian}, F., {Mezzetti}, M., \& {Bertotti}, G. 1990,
  \apjs, 72, 551

\bibitem[{{Glanz}(1992)}]{Glanz1992}
{Glanz}, S.~A. 1992, {Primer of Biostatistics} (New York: McGraw-Hill, 3rd Ed.)

\bibitem[{{Greenhill} \& {Gwinn}(1997)}]{1997Ap&SS.248..261G}
{Greenhill}, L.~J. \& {Gwinn}, C.~R. 1997, \apss, 248, 261

\bibitem[{{Griffith} {et~al.}(1995){Griffith}, {Wright}, {Burke}, \&
  {Ekers}}]{1995ApJS...97..347G}
{Griffith}, M.~R., {Wright}, A.~E., {Burke}, B.~F., \& {Ekers}, R.~D. 1995,
  \apjs, 97, 347

\bibitem[{{Hagiwara} {et~al.}(2000){Hagiwara}, {Diamond}, {Nakai}, \&
  {Kawabe}}]{2000AaA...360...49H}
{Hagiwara}, Y., {Diamond}, P.~J., {Nakai}, N., \& {Kawabe}, R. 2000, \aap, 360,
  49

\bibitem[{{Hagiwara} {et~al.}(2001){Hagiwara}, {Diamond}, {Nakai}, \&
  {Kawabe}}]{2001ApJ...560..119H}
---. 2001, \apj, 560, 119

\bibitem[{{Harrison} {et~al.}(1986){Harrison}, {Pedlar}, {Unger}, {Burgess},
  {Graham}, \& {Preuss}}]{1986MNRAS.218..775H}
{Harrison}, B., {Pedlar}, A., {Unger}, S.~W., {et~al.} 1986, \mnras, 218, 775

\bibitem[{{Herrnstein} {et~al.}(1997){Herrnstein}, {Moran}, {Greenhill},
  {Diamond}, {Miyoshi}, {Nakai}, \& {Inoue}}]{1997ApJ...475L..17H}
{Herrnstein}, J.~R., {Moran}, J.~M., {Greenhill}, L.~J., {et~al.} 1997, \apjl,
  475, L17

\bibitem[{{Ho} \& {Ulvestad}(2001)}]{2001ApJS..133...77H}
{Ho}, L.~C. \& {Ulvestad}, J.~S. 2001, \apjs, 133, 77

\bibitem[{{Hutchings}(1996)}]{1996AJ....111..712H}
{Hutchings}, J.~B. 1996, \aj, 111, 712

\bibitem[{{Irwin} \& {Seaquist}(1988)}]{1988ApJ...335..658I}
{Irwin}, J.~A. \& {Seaquist}, E.~R. 1988, \apj, 335, 658

\bibitem[{{Ivezi{\' c}} {et~al.}(2002){Ivezi{\' c}}, {Menou}, {Knapp},
  {Strauss}, {Lupton}, {Vanden Berk}, {Richards}, {Tremonti}, {Weinstein},
  {Anderson}, {Bahcall}, {Becker}, {Bernardi}, {Blanton}, {Eisenstein}, {Fan},
  {Finkbeiner}, {Finlator}, {Frieman}, {Gunn}, {Hall}, {Kim}, {Kinkhabwala},
  {Narayanan}, {Rockosi}, {Schlegel}, {Schneider}, {Strateva}, {SubbaRao},
  {Thakar}, {Voges}, {White}, {Yanny}, {Brinkmann}, {Doi}, {Fukugita},
  {Hennessy}, {Munn}, {Nichol}, \& {York}}]{2002AJ....124.2364I}
{Ivezi{\' c}}, {\v Z}., {Menou}, K., {Knapp}, G.~R., {et~al.} 2002, \aj, 124,
  2364

\bibitem[{{Kadler} {et~al.}(2003){Kadler}, {Ros}, {Kerp}, {Falcke}, {Zensus},
  {Pogge}, \& {Bicknell}}]{Kadler2003}
{Kadler}, M., {Ros}, E., {Kerp}, J., {et~al.} 2003, in The Physics of
  Relativistic Jets in the CHANDRA and XMM Era, Eds.: G. Brunetti, D.E. Harris,
  R.M. Sambruna, G. Setti

\bibitem[{{Kadler} {et~al.}(2002){Kadler}, {Ros}, {Lobanov}, \&
  {Falcke}}]{Kadler2002}
{Kadler}, M., {Ros}, E., {Lobanov}, A.~P., \& {Falcke}, H. 2002, in Proceedings
  of SRT: the impact of large antennas on Radio Astronomy and Space Science

\bibitem[{{Kameno} {et~al.}(2001){Kameno}, {Sawada-Satoh}, {Inoue}, {Shen}, \&
  {Wajima}}]{2001PASJ...53..169K}
{Kameno}, S., {Sawada-Satoh}, S., {Inoue}, M., {Shen}, Z., \& {Wajima}, K.
  2001, \pasj, 53, 169

\bibitem[{{Keel}(1996)}]{1996ApJS..106...27K}
{Keel}, W.~C. 1996, \apjs, 106, 27

\bibitem[{{Kellermann} {et~al.}(1989){Kellermann}, {Sramek}, {Schmidt},
  {Shaffer}, \& {Green}}]{1989AJ.....98.1195K}
{Kellermann}, K.~I., {Sramek}, R., {Schmidt}, M., {Shaffer}, D.~B., \& {Green},
  R. 1989, \aj, 98, 1195

\bibitem[{{Kinney} {et~al.}(2000){Kinney}, {Schmitt}, {Clarke}, {Pringle},
  {Ulvestad}, \& {Antonucci}}]{2000ApJ...537..152K}
{Kinney}, A.~L., {Schmitt}, H.~R., {Clarke}, C.~J., {et~al.} 2000, \apj, 537,
  152

\bibitem[{{Kukula} {et~al.}(1999){Kukula}, {Ghosh}, {Pedlar}, \&
  {Schilizzi}}]{1999ApJ...518..117K}
{Kukula}, M.~J., {Ghosh}, T., {Pedlar}, A., \& {Schilizzi}, R.~T. 1999, \apj,
  518, 117

\bibitem[{{Kukula} {et~al.}(1993){Kukula}, {Ghosh}, {Pedlar}, {Schilizzi},
  {Miley}, {de Bruyn}, \& {Saikia}}]{1993MNRAS.264..893K}
{Kukula}, M.~J., {Ghosh}, T., {Pedlar}, A., {et~al.} 1993, \mnras, 264, 893

\bibitem[{{Laor}(2000)}]{2000ApJ...543L.111L}
{Laor}, A. 2000, \apjl, 543, L111

\bibitem[{{Lavalley} {et~al.}(1992){Lavalley}, {Isobe}, \&
  {Feigelson}}]{1992adass...1..245L}
{Lavalley}, M., {Isobe}, T., \& {Feigelson}, E. 1992, in ASP Conf. Ser. 25:
  Astronomical Data Analysis Software and Systems I, 245

\bibitem[{{Lestrade}(1991)}]{1991ritt.proc..289L}
{Lestrade}, J.-F. 1991, in ASP Conf. Ser. 19: IAU Colloq. 131: Radio
  Interferometry. Theory, Techniques, and Applications, 289--297

\bibitem[{{Martin} {et~al.}(1983){Martin}, {Thompson}, {Maza}, \&
  {Angel}}]{1983ApJ...266..470M}
{Martin}, P.~G., {Thompson}, I.~B., {Maza}, J., \& {Angel}, J.~R.~P. 1983,
  \apj, 266, 470

\bibitem[{{Meier} {et~al.}(1997){Meier}, {Edgington}, {Godon}, {Payne}, \&
  {Lind}}]{1997Natur.388..350M}
{Meier}, D.~L., {Edgington}, S., {Godon}, P., {Payne}, D.~G., \& {Lind}, K.~R.
  1997, \nat, 388, 350

\bibitem[{{Miller} {et~al.}(1990){Miller}, {Peacock}, \&
  {Mead}}]{1990MNRAS.244..207M}
{Miller}, L., {Peacock}, J.~A., \& {Mead}, A.~R.~G. 1990, \mnras, 244, 207

\bibitem[{{Momjian} {et~al.}(2003){Momjian}, Romney, Carilli, \&
  Troland}]{Momjian2003}
{Momjian}, E., Romney, J.~D., Carilli, C.~L., \& Troland, T.~H. 2003, ApJ,
  accepted, astro-ph/0307399

\bibitem[{{Morganti} {et~al.}(1998){Morganti}, {Oosterloo}, \&
  {Tsvetanov}}]{1998AJ....115..915M}
{Morganti}, R., {Oosterloo}, T., \& {Tsvetanov}, Z. 1998, \aj, 115, 915

\bibitem[{{Morganti} {et~al.}(1999){Morganti}, {Tsvetanov}, {Gallimore}, \&
  {Allen}}]{1999AAS..137..457M}
{Morganti}, R., {Tsvetanov}, Z.~I., {Gallimore}, J., \& {Allen}, M.~G. 1999,
  \aaps, 137, 457

\bibitem[{{Mundell} {et~al.}(1995){Mundell}, {Holloway}, {Pedlar}, {Meaburn},
  {Kukula}, \& {Axon}}]{1995MNRAS.275...67M}
{Mundell}, C.~G., {Holloway}, A.~J., {Pedlar}, A., {et~al.} 1995, \mnras, 275,
  67

\bibitem[{{Mundell} {et~al.}(2000){Mundell}, {Wilson}, {Ulvestad}, \&
  {Roy}}]{2000ApJ...529..816M}
{Mundell}, C.~G., {Wilson}, A.~S., {Ulvestad}, J.~S., \& {Roy}, A.~L. 2000,
  \apj, 529, 816

\bibitem[{{Murray} {et~al.}(1999){Murray}, {Gallimore}, {Baum}, {Pedlar},
  {Thean}, \& {Kukula}}]{Murray1999}
{Murray}, C., {Gallimore}, J.~F., {Baum}, S.~A., {et~al.} 1999, Bulletin of the
  American Astronomical Society, 31, 1544

\bibitem[{{Nagar} {et~al.}(2002{\natexlab{a}}){Nagar}, {Falcke}, {Wilson}, \&
  {Ulvestad}}]{2002AaA...392...53N}
{Nagar}, N.~M., {Falcke}, H., {Wilson}, A.~S., \& {Ulvestad}, J.~S.
  2002{\natexlab{a}}, \aap, 392, 53

\bibitem[{{Nagar} {et~al.}(2002{\natexlab{b}}){Nagar}, {Oliva}, {Marconi}, \&
  {Maiolino}}]{2002AaA...391L..21N}
{Nagar}, N.~M., {Oliva}, E., {Marconi}, A., \& {Maiolino}, R.
  2002{\natexlab{b}}, \aap, 391, L21

\bibitem[{{Nagar} \& {Wilson}(1999)}]{1999ApJ...516...97N}
{Nagar}, N.~M. \& {Wilson}, A.~S. 1999, \apj, 516, 97

\bibitem[{{Nagar} {et~al.}(1999){Nagar}, {Wilson}, {Mulchaey}, \&
  {Gallimore}}]{1999ApJS..120..209N}
{Nagar}, N.~M., {Wilson}, A.~S., {Mulchaey}, J.~S., \& {Gallimore}, J.~F. 1999,
  \apjs, 120, 209

\bibitem[{{Neff} \& {de Bruyn}(1983)}]{1983AaA...128..318N}
{Neff}, S.~G. \& {de Bruyn}, A.~G. 1983, \aap, 128, 318

\bibitem[{{Nelson} \& {Whittle}(1995)}]{1995ApJS...99...67N}
{Nelson}, C.~H. \& {Whittle}, M. 1995, \apjs, 99, 67

\bibitem[{{Nishiura} {et~al.}(2000){Nishiura}, {Shimada}, {Ohyama}, {Murayama},
  \& {Taniguchi}}]{2000AJ....120.1691N}
{Nishiura}, S., {Shimada}, M., {Ohyama}, Y., {Murayama}, T., \& {Taniguchi}, Y.
  2000, \aj, 120, 1691

\bibitem[{{Norman} \& {Miley}(1984)}]{1984AaA...141...85N}
{Norman}, C. \& {Miley}, G. 1984, \aap, 141, 85

\bibitem[{{Oosterloo} {et~al.}(2000){Oosterloo}, {Morganti}, {Tzioumis},
  {Reynolds}, {King}, {McCulloch}, \& {Tsvetanov}}]{2000AJ....119.2085O}
{Oosterloo}, T.~A., {Morganti}, R., {Tzioumis}, A., {et~al.} 2000, \aj, 119,
  2085

\bibitem[{{Peacock} {et~al.}(1986){Peacock}, {Miller}, \&
  {Longair}}]{1986MNRAS.218..265P}
{Peacock}, J.~A., {Miller}, L., \& {Longair}, M.~S. 1986, \mnras, 218, 265

\bibitem[{{Pearson} \& {Readhead}(1988)}]{1988ApJ...328..114P}
{Pearson}, T.~J. \& {Readhead}, A.~C.~S. 1988, \apj, 328, 114

\bibitem[{{Peck} {et~al.}(2003){Peck}, {Henkel}, {Ulvestad}, {Brunthaler},
  {Falcke}, {Elitzur}, {Menten}, \& {Gallimore}}]{2003ApJ...590..149P}
{Peck}, A.~B., {Henkel}, C., {Ulvestad}, J.~S., {et~al.} 2003, \apj, 590, 149

\bibitem[{{Pedlar} {et~al.}(1990){Pedlar}, {Ghataure}, {Davies}, {Harrison},
  {Perley}, {Crane}, \& {Unger}}]{1990MNRAS.246..477P}
{Pedlar}, A., {Ghataure}, H.~S., {Davies}, R.~D., {et~al.} 1990, \mnras, 246,
  477

\bibitem[{{Pedlar} {et~al.}(1993){Pedlar}, {Kukula}, {Longley}, {Muxlow},
  {Axon}, {Baum}, {O'Dea}, \& {Unger}}]{1993MNRAS.263..471P}
{Pedlar}, A., {Kukula}, M.~J., {Longley}, D.~P.~T., {et~al.} 1993, \mnras, 263,
  471

\bibitem[{{Pedlar} {et~al.}(1985){Pedlar}, {Unger}, \&
  {Dyson}}]{1985MNRAS.214..463P}
{Pedlar}, A., {Unger}, S.~W., \& {Dyson}, J.~E. 1985, \mnras, 214, 463

\bibitem[{{Polatidis} {et~al.}(2002){Polatidis}, {Conway}, \&
  {Owsianik}}]{Polatidis2002}
{Polatidis}, A.~G., {Conway}, J.~E., \& {Owsianik}, I. 2002, in Proceedings of
  the 6th European VLBI Network Symposium, Eds.: E. Ros, R. W. Porcas, A.P.
  Lobanov, \& J.A. Zensus, (Bonn: MPIfR), 139

\bibitem[{{Pringle}(1997)}]{1997MNRAS.292..136P}
{Pringle}, J. 1997, \mnras, 292, 136

\bibitem[{{Pringle} {et~al.}(1999){Pringle}, {Antonucci}, {Clarke}, {Kinney},
  {Schmitt}, \& {Ulvestad}}]{1999ApJ...526L...9P}
{Pringle}, J.~E., {Antonucci}, R.~R.~J., {Clarke}, C.~J., {et~al.} 1999, \apjl,
  526, L9

\bibitem[{{Roy} {et~al.}(1998){Roy}, {Colbert}, {Wilson}, \&
  {Ulvestad}}]{1998ApJ...504..147R}
{Roy}, A.~L., {Colbert}, E.~J.~M., {Wilson}, A.~S., \& {Ulvestad}, J.~S. 1998,
  \apj, 504, 147

\bibitem[{{Roy} {et~al.}(2000){Roy}, {Wilson}, {Ulvestad}, \&
  {Colbert}}]{2000evn..proc....7R}
{Roy}, A.~L., {Wilson}, A.~S., {Ulvestad}, J.~S., \& {Colbert}, J.~M. 2000, in
  EVN Symposium 2000, Proceedings of the 5th european VLBI Network Symposium
  held at Chalmers University of Technology, Gothenburg, Sweden, June 29 - July
  1, 2000, Eds.: J.E. Conway, A.G. Polatidis, R.S. Booth and Y.M. Pihlstr{\"
  o}m, published Onsala Space Observatory, p. 7

\bibitem[{{Rush} {et~al.}(1996){Rush}, {Malkan}, \&
  {Edelson}}]{1996ApJ...473..130R}
{Rush}, B., {Malkan}, M.~A., \& {Edelson}, R.~A. 1996, \apj, 473, 130

\bibitem[{{Sanghera} {et~al.}(1995){Sanghera}, {Saikia}, {Luedke}, {Spencer},
  {Foulsham}, {Akujor}, \& {Tzioumis}}]{1995AaA...295..629S}
{Sanghera}, H.~S., {Saikia}, D.~J., {Luedke}, E., {et~al.} 1995, \aap, 295, 629

\bibitem[{{Sawada-Satoh} {et~al.}(2000){Sawada-Satoh}, {Inoue}, {Shibata},
  {Kameno}, {Migenes}, {Nakai}, \& {Diamond}}]{2000PASJ...52..421S}
{Sawada-Satoh}, S., {Inoue}, M., {Shibata}, K.~M., {et~al.} 2000, \pasj, 52,
  421

\bibitem[{{Schmitt} \& {Kinney}(2000)}]{Schmitt2000}
{Schmitt}, H.~R. \& {Kinney}, A.~L. 2000, \apjs, 128, 479

\bibitem[{{Schmitt} {et~al.}(2002){Schmitt}, {Ulvestad}, \&
  {Antonucci}}]{2002AAS...200.4507S}
{Schmitt}, H.~R., {Ulvestad}, J.~S., \& {Antonucci}, R.~R.~J. 2002, American
  Astronomical Society Meeting, 34, 715

\bibitem[{{Storchi-Bergmann} {et~al.}(1998){Storchi-Bergmann}, {Fernandes}, \&
  {Schmitt}}]{1998ApJ...501...94S}
{Storchi-Bergmann}, T., {Fernandes}, R.~C., \& {Schmitt}, H.~R. 1998, \apj,
  501, 94

\bibitem[{{Su} {et~al.}(1996){Su}, {Muxlow}, {Pedlar}, {Holloway}, {Steffen},
  {Kukula}, \& {Mutel}}]{1996MNRAS.279.1111S}
{Su}, B.~M., {Muxlow}, T.~W.~B., {Pedlar}, A., {et~al.} 1996, \mnras, 279, 1111

\bibitem[{{Taylor} {et~al.}(1989){Taylor}, {Dyson}, {Axon}, \&
  {Pedlar}}]{1989MNRAS.240..487T}
{Taylor}, D., {Dyson}, J.~E., {Axon}, D.~J., \& {Pedlar}, A. 1989, \mnras, 240,
  487

\bibitem[{{Tran}(1995)}]{1995ApJ...440..578T}
{Tran}, H.~D. 1995, \apj, 440, 578

\bibitem[{{Tran} {et~al.}(1992){Tran}, {Miller}, \&
  {Kay}}]{1992ApJ...397..452T}
{Tran}, H.~D., {Miller}, J.~S., \& {Kay}, L.~E. 1992, \apj, 397, 452

\bibitem[{{Trotter} {et~al.}(1998){Trotter}, {Greenhill}, {Moran}, {Reid},
  {Irwin}, \& {Lo}}]{1998ApJ...495..740T}
{Trotter}, A.~S., {Greenhill}, L.~J., {Moran}, J.~M., {et~al.} 1998, \apj, 495,
  740

\bibitem[{{Ulvestad} \& {Ho}(2001{\natexlab{a}})}]{2001ApJ...558..561U}
{Ulvestad}, J.~S. \& {Ho}, L.~C. 2001{\natexlab{a}}, \apj, 558, 561

\bibitem[{{Ulvestad} \& {Ho}(2001{\natexlab{b}})}]{2001ApJ...562L.133U}
---. 2001{\natexlab{b}}, \apjl, 562, L133

\bibitem[{{Ulvestad} {et~al.}(1987){Ulvestad}, {Neff}, \&
  {Wilson}}]{1987AJ.....93...22U}
{Ulvestad}, J.~S., {Neff}, S.~G., \& {Wilson}, A.~S. 1987, \aj, 93, 22

\bibitem[{{Ulvestad} {et~al.}(1998){Ulvestad}, {Roy}, {Colbert}, \&
  {Wilson}}]{1998ApJ...496..196U}
{Ulvestad}, J.~S., {Roy}, A.~L., {Colbert}, E.~J.~M., \& {Wilson}, A.~S. 1998,
  \apj, 496, 196

\bibitem[{{Ulvestad} \& {Wilson}(1983)}]{1983ApJ...264L...7U}
{Ulvestad}, J.~S. \& {Wilson}, A.~S. 1983, \apjl, 264, L7

\bibitem[{{Ulvestad} \& {Wilson}(1984)}]{1984ApJ...285..439U}
---. 1984, \apj, 285, 439

\bibitem[{{Ulvestad} \& {Wilson}(1989)}]{1989ApJ...343..659U}
---. 1989, \apj, 343, 659

\bibitem[{{Ulvestad} {et~al.}(2002){Ulvestad}, {Wong}, {Taylor}, {Mundell}, \&
  {Gallimore}}]{2002AAS...201.4814U}
{Ulvestad}, J.~S., {Wong}, D.~S., {Taylor}, G.~B., {Mundell}, C.~G., \&
  {Gallimore}, J.~W. 2002, American Astronomical Society Meeting, 201, 0

\bibitem[{{Ulvestad} {et~al.}(1999{\natexlab{a}}){Ulvestad}, {Wrobel}, \&
  {Carilli}}]{1999ApJ...516..127U}
{Ulvestad}, J.~S., {Wrobel}, J.~M., \& {Carilli}, C.~L. 1999{\natexlab{a}},
  \apj, 516, 127

\bibitem[{{Ulvestad} {et~al.}(1999{\natexlab{b}}){Ulvestad}, {Wrobel}, {Roy},
  {Wilson}, {Falcke}, \& {Krichbaum}}]{1999ApJ...517L..81U}
{Ulvestad}, J.~S., {Wrobel}, J.~M., {Roy}, A.~L., {et~al.} 1999{\natexlab{b}},
  \apjl, 517, L81

\bibitem[{{Unger} {et~al.}(1987){Unger}, {Lawrence}, {Wilson}, {Elvis}, \&
  {Wright}}]{1987MNRAS.228..521U}
{Unger}, S.~W., {Lawrence}, A., {Wilson}, A.~S., {Elvis}, M., \& {Wright},
  A.~E. 1987, \mnras, 228, 521

\bibitem[{{Unger} {et~al.}(1988){Unger}, {Pedlar}, {Axon}, {Graham},
  {Harrison}, {Saikia}, {Whittle}, {Meurs}, {Dyson}, \&
  {Taylor}}]{1988MNRAS.234..745U}
{Unger}, S.~W., {Pedlar}, A., {Axon}, D.~J., {et~al.} 1988, \mnras, 234, 745

\bibitem[{{Unger} {et~al.}(1986){Unger}, {Pedlar}, {Booler}, \&
  {Harrison}}]{1986MNRAS.219..387U}
{Unger}, S.~W., {Pedlar}, A., {Booler}, R.~V., \& {Harrison}, B.~A. 1986,
  \mnras, 219, 387

\bibitem[{{V{\' e}ron-Cetty} \& {V{\' e}ron}(1991)}]{1991cqan.book.....V}
{V{\' e}ron-Cetty}, M.-P. \& {V{\' e}ron}, P. 1991, {A Catalogue of quasars and
  active nuclei} (ESO Scientific Report, Garching: European Southern
  Observatory (ESO), 1991, 5th ed.)

\bibitem[{{V{\' e}ron-Cetty} \& {V{\' e}ron}(1998)}]{1998csan.book.....V}
---. 1998, {A Catalogue of quasars and active nuclei} (A Catalogue of quasars
  and active nuclei, Edition: 8th ed., Publisher: Garching: European Southern
  Observatory (ESO), 1998, Series: ESO Scientific Report Series vol no: 18)

\bibitem[{{V{\' e}ron-Cetty} \& {V{\' e}ron}(2001)}]{Veron2001}
---. 2001, \aap, 374, 92

\bibitem[{{Vermeulen} {et~al.}(2003){Vermeulen}, {Ros}, {Kellermann}, {Cohen},
  {Zensus}, \& {van Langevelde}}]{2003AaA...401..113V}
{Vermeulen}, R.~C., {Ros}, E., {Kellermann}, K.~I., {et~al.} 2003, \aap, 401,
  113

\bibitem[{{Vir Lal}(2001)}]{VirLal2001}
{Vir Lal}, V. 2001, Ph.D.~Thesis, Indian Institute of Astrophysics, Bangalore,
  India

\bibitem[{{Walker} {et~al.}(2000){Walker}, {Dhawan}, {Romney}, {Kellermann}, \&
  {Vermeulen}}]{2000ApJ...530..233W}
{Walker}, R.~C., {Dhawan}, V., {Romney}, J.~D., {Kellermann}, K.~I., \&
  {Vermeulen}, R.~C. 2000, \apj, 530, 233

\bibitem[{{White} {et~al.}(2000){White}, {Becker}, {Gregg},
  {Laurent-Muehleisen}, {Brotherton}, {Impey}, {Petry}, {Foltz}, {Chaffee},
  {Richards}, {Oegerle}, {Helfand}, {McMahon}, \&
  {Cabanela}}]{2000ApJS..126..133W}
{White}, R.~L., {Becker}, R.~H., {Gregg}, M.~D., {et~al.} 2000, \apjs, 126, 133

\bibitem[{{Whittle}(1992)}]{1992ApJ...387..121W}
{Whittle}, M. 1992, \apj, 387, 121

\bibitem[{{Whittle} {et~al.}(1986){Whittle}, {Haniff}, {Ward}, {Meurs},
  {Pedlar}, {Unger}, {Axon}, \& {Harrison}}]{1986MNRAS.222..189W}
{Whittle}, M., {Haniff}, C.~A., {Ward}, M.~J., {et~al.} 1986, \mnras, 222, 189

\bibitem[{{Whittle} {et~al.}(1988){Whittle}, {Pedlar}, {Meurs}, {Unger},
  {Axon}, \& {Ward}}]{1988ApJ...326..125W}
{Whittle}, M., {Pedlar}, A., {Meurs}, E.~J.~A., {et~al.} 1988, \apj, 326, 125

\bibitem[{{Wilson} \& {Colbert}(1995)}]{1995ApJ...438...62W}
{Wilson}, A.~S. \& {Colbert}, E.~J.~M. 1995, \apj, 438, 62

\bibitem[{{Wilson} {et~al.}(1976){Wilson}, {Penston}, {Fosbury}, \&
  {Boksenberg}}]{1976MNRAS.177..673W}
{Wilson}, A.~S., {Penston}, M.~V., {Fosbury}, R.~A.~E., \& {Boksenberg}, A.
  1976, \mnras, 177, 673

\bibitem[{{Wilson} {et~al.}(1998){Wilson}, {Roy}, {Ulvestad}, {Colbert},
  {Weaver}, {Braatz}, {Henkel}, {Matsuoka}, {Xue}, {Iyomoto}, \&
  {Okada}}]{1998ApJ...505..587W}
{Wilson}, A.~S., {Roy}, A.~L., {Ulvestad}, J.~S., {et~al.} 1998, \apj, 505, 587

\bibitem[{{Wilson} \& {Tsvetanov}(1994)}]{1994AJ....107.1227W}
{Wilson}, A.~S. \& {Tsvetanov}, Z.~I. 1994, \aj, 107, 1227

\bibitem[{{Wrobel}(2000)}]{2000ApJ...531..716W}
{Wrobel}, J.~M. 2000, \apj, 531, 716

\bibitem[{{Wrobel} {et~al.}(2001){Wrobel}, {Fassnacht}, \&
  {Ho}}]{2001ApJ...553L..23W}
{Wrobel}, J.~M., {Fassnacht}, C.~D., \& {Ho}, L.~C. 2001, \apjl, 553, L23

\end{thebibliography}

\begin{table*}
\tiny
\begin{center}
\begin{tabular}{lrrc.cc}
\hline
\hline
& \multicolumn{1}{c}{Int. flux density} 
& \multicolumn{1}{c}{Peak flux density} 
& \multicolumn{1}{c}{Peak position (J2000)}
& \multicolumn{1}{c}{rms} 
& \multicolumn{1}{c}{Beam size} 
& \multicolumn{1}{c}{$\alpha^{18}_6$}\\

& \multicolumn{1}{c}{(mJy)} 
& \multicolumn{1}{c}{(mJy/beam)} 
&  \multicolumn{1}{c}{RA~~~~~~~~~~~~~~~~~~Dec} 
& \multicolumn{1}{c}{(mJy/beam)} 
& \multicolumn{1}{c}{(mas)} 
& ($S\propto \nu^{\alpha}$)\\
\hline
\\
&&&\multicolumn{1}{c}{\bf \object{NGC~5506}}\vspace{0.1cm}\\
{\bf global VLBI, 6~cm, 28 Feb 1994}\\
B0       & $45.7\pm5.9$    & $33.9\pm4.3$    & 0.0~~0.0    &  0.9 & 12.6$\times$6.0\\
         &                 &                 &\\
B1       & $30.0\pm4.2$    & $12.2\pm2.1$    & 28.0~mas~~7.3~mas\\
         &                 &                 &\\
B2       & $10.6\pm2.5$     & $4.9\pm1.4$    & 41.6~mas~~$-$8.3~mas\\
         &                 &                 & \\
\\
{\bf VLBA 1997, 18~cm, Feb 5, 1997}\\
B0       & $27.9\pm3.1$    & $24.0\pm2.7$   & 0.0~~0.0    & 0.26    & $18.4\times11.4$\\
B1       & $24.4\pm2.9$    & $15.9\pm1.9$   & 29.7 mas~~9.7 mas\\
B2       & $6.8\pm1.3$     & $4.0\pm0.7$    & 37.1 mas~~$-$10.5 mas\\
\\
{\bf VLBA 1997, 6~cm, Feb 5, 1997}\\
B0       & $41.6\pm4.9$    & $29.6\pm3.3$   & 0.0~~0.0    & 0.30    & $5.7\times3.7$\\
B1       & $24.4\pm3.0$    & $4.3\pm0.7$    & 28.9 mas~~9.1 mas\\
\\
{\bf VLBA 1997, 3.6~cm, Feb 5, 1997}\\
B0       & $27.0\pm3.4$    & $21.5\pm2.6$   & 0.0~~0.0    & 0.46    & $6.5\times3.0$\\
\\
{\bf EVN, 18~cm, Nov 11, 1999}\\
B0      & $32.9\pm4.1$   & $31.0\pm3.5$            & 0.0~~0.0                        & 0.43      & 26.2$\times$14.1 & 0.06\\
B1      & $25.6\pm3.3$   & $22.1\pm2.6$            & 26.2 mas~~9.3 mas               &           &                  & $-$0.73\\
B2      & $ 6.1\pm1.4$   & $5.4\pm1.0$             & 37.9 mas~~$-$15.2 mas             &           &                  & $-$0.30\\
\\
{\bf MERLIN, 18~cm, Nov 11, 1999}\\
 & $113.2\pm10.3$   & $86.6\pm6.7$                  & 0.0~~0.0 & 2.41      & 259.1$\times$150.9 & $-$0.33\\
\\
{\bf  EVN, 6~cm, Feb 28, 2000}\\
B0       & $35.9\pm4.1$   & $36.1\pm4.2$            & 14~13~14.87720~~$-$03~12~27.6460  & 0.20      & 8.8$\times$6.8\\
(tapered)& $32.0\pm3.7$\\
B1       & $11.0\pm1.4$    & $6.8\pm1.2$             & 14~13~14.87916~~$-$03~12~27.6365\\
(tapered)& $10.8\pm1.4$\\
B2       & $1.9\pm0.5$    & $1.3\pm0.4$             & 14~13~14.87982~~$-$03~12~27.6583\\
(tapered)& $4.4\pm0.7$\\
\\
{\bf MERLIN, 6~cm, Feb 28, 2000}\\
B0+B1+B2 & $85.8\pm6.4$    & $66.8\pm3.8$           & 0.0~~0.0 & 0.43      & 78.6$\times$62.4\\
(tapered)& $78.8\pm5.9$\\
\\
{\bf VLBA 2000, 18~cm, May 31, 2000}\\
B0       & $49.0\pm5.8$    & $46.0\pm5.1$   & 0.0~~0.0   & 0.45 & 18.6$\times$11.3\\
B1       & $35.8\pm4.4$    & $24.7\pm2.9$   & 29.7~mas~~10.1~mas\\
B2       & $4.6\pm1.1$\\
\\
{\bf VLBA 2000, 6~cm, May 31, 2000}\\
B0       & $48.1\pm5.2$    & $42.2\pm4.6$   & 0.0~~0.0   & 0.42 & $7.4\times4.3$\\
B1       & $15.3\pm2.7$    & $6.6\pm1.1$    & 29.1~mas~~9.6~mas\\
B2       & $2.3\pm0.8$     & $2.7\pm0.7$    & 45.2~mas~~$-$8.5~mas\\
\\
\hline
 & & & \multicolumn{1}{c}{{\bf \object{NGC~7674}}}\vspace{0.1cm}\\
{\bf EVN, 18~cm, Apr 13, 1985}\\
SE       &  $31.4\pm3.9$            &  $31.4\pm3.9$     & 0.0~~0.0 & 0.46 & $39.3\times24.2$\\
NW       &  $4.1\pm1.1$            &  $5.1\pm0.97$      & $-$453.1 mas~~230.8 mas\\
\\
{\bf  EVN, 18~cm, Nov 11, 1999}\\
SE      & $37.6\pm4.5$ & $23.9\pm2.8$   & 23~27~56.71054~~+08~46~44.1310 & 0.40 & 32.5$\times$9.8 & $-$1.78\\
NW      & $11.0\pm2.1$ & $5.6\pm1.0$    & 23~27~56.67984~~+08~46~44.3641\\
\\
{\bf  MERLIN, 18~cm, Nov 11, 1999}\\
SE      & $111.5\pm8.6$ & $74.9\pm4.6$  & 23~27~56.71730~~+08~46~44.0104 & 0.82 & 276$\times$110 & $-$1.45\\
NW      & $45.6\pm4.0$  & $29.2\pm2.3$  & 23~27~56.68608~~+08~46~44.2195 & & & $-$1.58\\
\\
{\bf  EVN, 6~cm, Feb 28, 2000}\\
SE       & $6.2\pm1.2$ & $5.0\pm0.7$    & 23~27~56.71201~~+08~46~44.1368 & 0.24 & 8.2$\times$4.6\\
(tapered)& $5.2\pm1.0$\\
\\
{\bf MERLIN, 6~cm, Feb 28, 2000}\\
SE       & $18.0\pm2.3$  & $16.2\pm1.1$  & 0.0~~0.0 & 0.31 & 82.2$\times$36.8\\
(tapered)& $18.7\pm2.4$\\
NW       & $3.8\pm0.7$  &  $3.9\pm0.5$  & $-$452.5 mas~~229.5 mas\\
(tapered)& $6.9\pm1.1$\\
\\
\\
\end{tabular}
\normalsize
\caption{Results from the observations.  ``Tapered'' 6~cm flux
densities were measured using images tapered and restored with the
18~cm beam size. The MERLIN 18~cm and 6~cm positions differ by 35~mas
because the coordinates used at the correlator were changed in an
unknown way and the resulting offset could not be removed
accurately. We have precessed the MERLIN B1950 coordinates to J2000
using NRAO's scheduling software SCHED. Positions are absolute for
observations that were phase-referenced and the phase-referencing
worked; such positions are given in h, m, s (for RA) and $^{\circ}$,
$^{\prime}$, $^{\prime\prime}$ (for Dec). Otherwise the positions are
given in mas relative to the strongest component. The flux density
error estimates represent $1~\sigma$.}
\label{tab:results}
\end{center}
\end{table*}

\newpage

\begin{table*}
\tiny
\begin{center}
\begin{tabular}{lrrc.cc}
\hline
\hline
& \multicolumn{1}{c}{Int. flux density} 
& \multicolumn{1}{c}{Peak flux density} 
& \multicolumn{1}{c}{Peak position (J2000)}
& \multicolumn{1}{c}{rms} 
& \multicolumn{1}{c}{Beam size} 
& \multicolumn{1}{c}{$\alpha^{18}_6$}\\

& \multicolumn{1}{c}{(mJy)} 
& \multicolumn{1}{c}{(mJy/beam)} 
& \multicolumn{1}{c}{RA~~~~~~~~~~~~~~~~~~Dec}
& \multicolumn{1}{c}{(mJy/beam)} 
& \multicolumn{1}{c}{(mas)} 
& ($S\propto \nu^{\alpha}$)\\
\hline
\\
&&&\multicolumn{1}{c}{\bf \object{NGC~2110}}\vspace{0.1cm}\\
{\bf EVN, 18~cm, Nov 11, 1999}\\
Core  & $15.5\pm1.1$      & $13.9\pm1.7$            & 05~52~11.37378~~$-$07~27~22.7071  & 0.32      & 26.2$\times$14.6 & 0.12\\
\\
{\bf MERLIN, 18~cm, Nov 11, 1999}\\
(tapered)\\
Core  & $29.4\pm2.9$   & $31.4\pm2.5$             & 05~52~11.38263~~$-$07~27~22.5310  & 0.89      & 318.0$\times$193.4 & 0.20\\
North & $11.4\pm2.1$   & $12.8\pm2.2$             & 05~52~11.39579~~$-$07~27~20.8662\\
South &  $9.1\pm1.9$   & $10.8\pm2.0$            & 05~52~11.37443~~$-$07~27~24.1280\\
\\
{\bf EVN, 6~cm, Feb 28, 2000}\\
Core     &  $20.3\pm2.7$      & $20.0\pm2.5$    & 0.0~~0.0 & 0.48      & 7.9$\times$6.4\\
(tapered)&  $17.6\pm2.4$\\\\
\\
{\bf MERLIN, 6~cm, Feb 28, 2000}\\
Core     & $36.6\pm2.9$      & $38.9\pm2.3$            & 0.0~~0.0  & 0.33        & 90.0$\times$67.7\\
\\
\hline
\\
&&&\multicolumn{1}{c}{\bf \object{Mrk~1210}}\vspace{0.1cm}\\
{\bf VLBA, 18~cm, Aug 6, 1998}\\
(natural weighting)\\
NW               & $37.4\pm3.8$ & $21.6\pm2.5$    &  08~04~05.85528~~+05~06~49.8530 & 0.16 & 18.6$\times$14.2\\
SE               & $11.8\pm1.9$ & $10.9\pm1.8$    &  08~04~05.85690~~+05~06~49.8292 &      &                 \\
\\
(uniform weighting)
NW               & $36.0\pm3.8$ & $12.8\pm1.6$    &  08~04~05.85514~~+05~06~49.8532 & 0.19 & 12.2$\times$8.6  & $-$1.26\\
SE               & $12.1\pm1.9$ & $8.5\pm1.5$     &  08~04~05.85701~~+05~06~49.8291 &      &                  & $-$0.78\\
\\
{\bf VLBA, 6~cm, Aug 6, 1998}\\
(tapered)\\
NW               & $9.1\pm1.1$  & $6.3\pm1.2$     &  08~04~05.85529~~+05~06~49.8552 & 0.31 & 12.3$\times$8.4\\
SE               & $5.0\pm1.1$  & $4.5\pm1.0$     &  08~04~05.85694~~+05~06~49.8304\\
\\
(full resolution)\\
NW1              & $1.0\pm0.2$   & $1.1\pm0.1$      &  08~04~05.85566~~+05~06~49.8570 & 0.12 & $3.2\times1.5$\\
NW2              & $2.1\pm0.4$   & $1.7\pm0.2$      &  08~04~05.85538~~+05~06~49.8554\\
NW3$^{1)}$       & $1.1\pm0.2$   & $0.9\pm0.1$      &  08~04~05.85511~~+05~06~49.8553\\
NW4$^{1)}$       & $1.2\pm0.3$   & $1.0\pm0.1$      &  08~04~05.85500~~+05~06~49.8557\\
SE               & $3.9\pm0.6$   & $2.6\pm0.3$      &  08~04~05.85696~~+05~06~49.8298\\
\\
{\bf EVN, 18~cm, Nov 11, 1999}\\
NW     & $36.6\pm4.3$   & $19.8\pm2.3$            & 08~04~05.85520~~+05~06~49.8545  & 0.33      & 24.6$\times$11.3 & $-$1.19\\
SE     & $13.0\pm2.3$   & $8.0\pm1.1$             & 08~04~05.85666~~+05~06~49.8301\\
\\
{\bf MERLIN, 18~cm, Nov 11, 1999}\\
       & $121.3\pm9.0$  & $99.3\pm5.5$            & 08~04~05.86024~~+05~06~49.8985  & 0.50      & 240.7$\times$136.5 & $-$1.01\\
\\
{\bf EVN, 6~cm, Feb 28, 2000}\\
       &  $9.8\pm2.1$  & $7.7\pm1.4$             & 0.0~~0.0  & 0.63      & 38.4$\times$28.0\\
\\
{\bf MERLIN, 6~cm, Feb 28, 2000}\\
         & $33.1\pm2.5$   & $24.1\pm1.4$            & 0.0~~0.0  & 0.21      & 72.7$\times$47.6\\
(tapered)& $39.4\pm3.0$\\
\\
\hline
\\
\end{tabular}
\normalsize
\caption{(continued) $^{1)}$ Positions and flux densities
measured by fitting Gaussians to the image, instead of fitting a
quadratic surface and integrating over the component region.}
\end{center}
\end{table*}

\begin{table*}
\begin{center}
\begin{tabular}{lllrll}
\hline
\hline
\multicolumn{1}{c}{Name} 
& \multicolumn{1}{c}{V\'eron name} 
& \multicolumn{1}{c}{Class}
& \multicolumn{1}{c}{Lin. Res.}
& \multicolumn{1}{c}{Instrument}
& \multicolumn{1}{c}{Ref.}
\\
&
&
& \multicolumn{1}{c}{(pc)}\\
\hline						  
\object{3C 287.1}    &                 & S1   & 5.43      & VLBA 5 GHz                    & \cite{2000ApJS..131...95F}\\  
\object{3C 390.3}    &                 & S1.5 & 1.09      & global VLBI 5 GHz             & \cite{1996AaA...308..376A}\\  
\object{III Zw 2}    &                 & S1.2 & 0.260     & VLBA 15/43 GHz                & \cite{2000AaA...357L..45B}\\  
\object{Ark 564}     &                 & -    & 0.598     & global VLBI 5 GHz             & \cite{VirLal2001}         \\  
\object{IC 5063}     & PKS 2048-57     & S1h  & 3.30      & LBA 2.3 GHz + LBA HI          & \cite{2000AJ....119.2085O}\\  
\object{MCG 8-11-11} &                 & S1.5 & 0.635     & global VLBI 5 GHz             & \cite{VirLal2001}         \\  
\object{Mrk 1}       &                 & S2   & 6.18      & EVN  1.7 GHz                  & \cite{1999ApJ...518..117K}\\  
\object{Mrk 1}       &                 & S2   & 0.550     & global VLBI 5 GHz             & \cite{VirLal2001}         \\  
\object{Mrk 3}       &                 & S1h  & 5.24      & EVN  1.7 GHz                  & \cite{1999ApJ...518..117K}\\  
\object{Mrk 78}      &                 & S2   & 1.75      & global VLBI 5 GHz             & \cite{VirLal2001}         \\  
\object{Mrk 231}     &                 & S1.0 & 0.343     & VLBA 15 GHz                   & \cite{1999ApJ...517L..81U}\\  
\object{Mrk 231}     &                 & S1.0 & 4.09      & VLBA 1.4/2.3/4.8/8.4/15/22 GHz& \cite{1999ApJ...516..127U}\\  
\object{Mrk 273}     &                 & S2   & 7.32      & VLBA 1.4 GHz                  & \cite{2000ApJ...532L..95C}\\  
\object{Mrk 348}     &                 & S1h  & 0.131     & VLBA 15 GHz                   & \cite{1999ApJ...517L..81U}\\  
\object{Mrk 348}     &                 & S1h  & 6.41      & EVN 1.4 GHz                   & \cite{1983AaA...128..318N}\\  
\object{Mrk 348}     &                 & S1h  & 1.49      & VLBA 1.7/5.0/8/15/22 GHz      & \cite{2003ApJ...590..149P}\\  
\object{Mrk 477}     &                 & S1h  & 2.39      & global VLBI 5 GHz             & \cite{VirLal2001}         \\  
\object{Mrk 530}     & NGC 7603        & S1.5 & 1.45      & global VLBI 5 GHz             & \cite{VirLal2001}         \\  
\object{Mrk 766}     &                 & S1.5 & 0.413     & global VLBI 5 GHz             & \cite{VirLal2001}         \\  
\object{Mrk 926}     &                 & S1.5 & 0.917     & VLBA 8.4 GHz                  & \cite{2000ApJ...529..816M}\\  
\object{Mrk 1210}    &                 & S1h  & 0.392     & EVN 1.6/5 GHz VLBA 1.6/5 GHz  & This paper                \\  
\object{Mrk 1218}    &                 & S1.8 & 0.582     & global VLBI 5 GHz             & \cite{VirLal2001}         \\  
\object{NGC 1052}    &                 & S3h  & 0.015     & VLBA 5/8.4/22/43 GHz          & \cite{Kadler2002}         \\  
\object{NGC 1052}    &                 & S3h  & 0.048     & VLBI 1.4-43.2                 & \cite{2003AaA...401..113V}\\  
\object{NGC 1052}    &                 & S3h  & 0.038     & VLBA 2.3/8.4/15.4 GHZ         & \cite{2001PASJ...53..169K}\\  
\object{NGC 1068}    &                 & S1h  & 0.221     & VLBA 1.7,5,15 GHz             & \cite{1998ApJ...504..147R}\\  
\object{NGC 1068}    &                 & S1h  & 2.94      & EVN 1.4 GHz                   & \cite{1987AJ.....93...22U}\\  
\object{NGC 1167}    &                 & S3   & 0.895     & global VLBI 5 GHz             & \cite{2001ApJ...552..508G}\\  
\object{NGC 1167}    &                 & S3   & 0.639     & global VLBI 1.6 GHz           & \cite{1990ApJ...358..159G}\\  
\object{NGC 1275}    &                 & S1.5 & 0.416     & VLBA 2.3/5/8.4/15.4/22/43 GHz & \cite{2000ApJ...530..233W}\\  
\object{NGC 2110}    &                 & S1i  & 0.151     & VLBA 8.4 GHz                  & \cite{2000ApJ...529..816M}\\  
\object{NGC 2110}    &                 & S1i  & 0.966     & EVN 1.6/5 GHz                 & This paper                \\  
\object{NGC 2273}    &                 &  -   & 0.342     & global VLBI 5 GHz             & \cite{VirLal2001}         \\  
\object{NGC 2639}    &                 & S3   & 0.233     & global VLBI 5 GHz             & \cite{VirLal2001}         \\  
\object{NGC 2639}    &                 & S3   & 0.162     & VLBA 1.6/5/15 GHz             & \cite{1998ApJ...505..587W}\\  
\object{NGC 3079}    &                 & S2   & 0.044     & VLBA 5/8/22 GHz               & \cite{1998ApJ...495..740T}\\  
\object{NGC 3079}    &                 & S2   & 0.054     & global VLBI, 5.0 GHz          & \cite{1988ApJ...335..658I}\\  
\object{NGC 3079}    &                 & S2   & 0.042     & global VLBI 1.4/8.4/15/22     & \cite{2000PASJ...52..421S}\\  
\object{NGC 3147}    &                 & S2   & 0.219     & VLBA 1.6/2.3/5/8.4 GHz        & \cite{2001ApJ...562L.133U}\\  
\object{NGC 3227}    &                 & S1.5 & 3.74      & MERLIN 1.6/5 GHz              & \cite{1995MNRAS.275...67M}\\  
\object{NGC 4151}    &                 & S1.5 & 0.116     & VLBA 1.6,5 GHz                & \cite{1998ApJ...496..196U}\\  
\object{NGC 4151}    &                 & S1.5 & 1.29      & EVN 1.7 GHz                   & \cite{1986MNRAS.218..775H}\\  
\object{NGC 4151}    &                 & S1.5 & 4.82      & MERLIN 5 GHz                  & \cite{1993MNRAS.263..471P}\\  
\object{NGC 4168}    &                 & S1.9 & 0.236     & VLBA 5 GHz                    & \cite{2002AaA...392...53N}\\  
\object{NGC 4203}    &                 & S3b  & 0.084     & VLBA 1.6/2.3/5/8.4 GHz        & \cite{2001ApJ...562L.133U}\\  
\object{NGC 4258}    &                 & S2   & 0.029     & VLBA 22 GHz                   & \cite{1997ApJ...475L..17H}\\  
\object{NGC 4258}    &                 & S2   & 0.232     & VLBA 1.4/1.6 GHz              & \cite{2000ApJ...536..675C}\\  
\object{NGC 4395}    &                 & S1.8 & 0.221     & VLBA 1.4 GHz                  & \cite{2001ApJ...553L..23W}\\  
\object{NGC 4565}    &                 & S1.9 & 0.207     & VLBA 5 GHz                    & \cite{2000ApJ...542..197F}\\  
\object{NGC 4579}    &                 & S3b  & 0.118     & VLBA 1.6/2.3/5/8.4 GHz        & \cite{2001ApJ...562L.133U}\\  
\object{NGC 4579}    &                 & S3b  & 0.246     & VLBA 5 GHz                    & \cite{2000ApJ...542..197F}\\  
\object{NGC 5252}    &                 & S2   & 0.671     & VLBA 8.4 GHz                  & \cite{2000ApJ...529..816M}\\  
\object{NGC 5506}    &                 & S1i  & 0.815     & EVN 1.6/5 GHz                 &      This paper           \\  
\object{NGC 5548}    &                 & S1.5 & 0.366     & VLBA 8.4 GHz                  & \cite{2000ApJ...531..716W}\\  
\object{NGC 5793}    &                 & S2   & 0.135     & VLBA 1.7,8.4,15,22 GHz        & \cite{2001ApJ...560..119H}\\  
\object{NGC 5793}    &                 & S2   & 0.316     & VLBA+Y27 1.6/5 GHz            & \cite{2000AaA...360...49H}\\  
\hline
\end{tabular}
\caption{}
\label{tab:seyferts}
\end{center}
\end{table*}

\begin{table*}
\begin{center}
\begin{tabular}{lllrll}
\multicolumn{6}{c}{(continued)}\\
\hline
\hline
\multicolumn{1}{c}{Name} 
& \multicolumn{1}{c}{V\'eron name} 
& \multicolumn{1}{c}{Class}
& \multicolumn{1}{c}{Lin. Res.}
& \multicolumn{1}{c}{Instrument}
& \multicolumn{1}{c}{Ref.}
\\
&
&
& \multicolumn{1}{c}{(pc)}\\
\hline						  
\object{NGC 5929}    &                 & S3   & 0.290     &  global VLBI 5 GHz            & \cite{VirLal2001}         \\  
\object{NGC 5929}    &                 & S3   & 6.44      & MERLIN 0.4/1.6/5              & \cite{1996MNRAS.279.1111S}\\  
\object{NGC 7212}    &                 & S1h  & 1.37      & global VLBI 5 GHz             & \cite{VirLal2001}         \\  
\object{NGC 7469}    &                 & S1.5 & 0.844     & global VLBI 5 GHz             & \cite{VirLal2001}         \\  
\object{NGC 7674}    &                 & S1h  & 2.80      & VLBA+Y27+Arecibo 1.4 GHz      & \cite{Momjian2003}        \\  
\object{NGC 7674}    &                 & S1h  & 2.58      & EVN 1.6/5 GHz                 & This paper                \\  
\object{NGC 7674}    &                 & S1h  & 2.18      & global VLBI 5 GHz             & \cite{VirLal2001}         \\  
\object{NGC 7682}    &                 & S2   & 0.996     & global VLBI 5 GHz             & \cite{VirLal2001}         \\  
\object{T0109-383}   & NGC  424        & S1h  & 0.226     & VLBA 8.4 GHz                  & \cite{2000ApJ...529..816M}\\  
\hline
\end{tabular}
\addtocounter{table}{-1}
\caption{All Seyferts observed with VLBI. Column 1 lists the names we
used, column 2 the names used by \cite{Veron2001}, column 3 the
Seyfert types as described by \cite{Veron2001} , column 4 the linear
resolution in pc, column 5 the instrument and frequency used, and
column 6 the reference to the publication.}
\end{center}
\end{table*}

\begin{table*}
\begin{center}
\begin{tabular}{lclrl}
\hline
\hline
\multicolumn{1}{c}{Source} 
& \multicolumn{1}{c}{Flat/Inv.}
& \multicolumn{1}{c}{Comp.}
& \multicolumn{1}{c}{$\alpha^{18}_{6}$}
& \multicolumn{1}{c}{Refs.}
\\
\hline
\object{Mrk  231}   & * &   N    &   $-$0.86  &   \cite{1999ApJ...516..127U}    \\ 
                    &   &   C    &   $+$0.88  &   \cite{1999ApJ...516..127U}    \\ 
                    &   &   S    &   $-$1.54  &   \cite{1999ApJ...516..127U}    \\ 
\object{Mrk  348}   & * &        &   $+$0.93  &   \cite{1998AJ....115..885B}    \\ 
\object{Mrk  463E}  &   &   L    &   $-$1.06  &   Norris et al., unpublished    \\ 
                    &   &   R    &   $-$0.62  &   Norris et al., unpublished    \\ 
                    &   &   1    &   $-$0.67  &   Norris et al., unpublished    \\ 
                    &   &   2    &   $-$0.77  &   Norris et al., unpublished    \\ 
                    &   &   3    &   $-$1.00  &   Norris et al., unpublished    \\ 
\object{Mrk 1210}   &   &   NW   &   $-$1.26  &   This paper                    \\ 
                    &   &   SE   &   $-$0.78  &   This paper                    \\ 
\object{NGC 1068}   & * &   NE   &   $-$1.5   &   \cite{1998ApJ...504..147R}    \\ 
                    &   &   C    &   $>2.2$   &   \cite{1998ApJ...504..147R}    \\ 
                    &   &   S1   &   $>0.6$   &   \cite{1998ApJ...504..147R}    \\ 
                    &   &   S2   &   $<-1.9$  &   \cite{1998ApJ...504..147R}    \\ 
\object{NGC 2110}   & * &   C    &   $+0.12$  &   This paper                    \\ 
\object{NGC 2639}   & * &        &   $+1.78$  &   \cite{1998ApJ...505..587W}    \\ 
\object{NGC 3079}   & * &   E    &   $+0.77$  &   Middelberg et al. in prep.    \\ 
                    &   &   A    &   $>+4.54$ &   Middelberg et al. in prep.    \\ 
                    &   &   B    &   $>+4.36$ &   Middelberg et al. in prep.    \\ 
\object{NGC 3147}   & * &        &   $+0.20$  &   \cite{2001ApJ...562L.133U}    \\ 
\object{NGC 4151}   & * &   E    &   $-0.5$   &   \cite{1998ApJ...496..196U}    \\ 
                    &   &   D    &   $<-0.2$  &   \cite{1998ApJ...496..196U}    \\ 
                    &   &   F    &   $<-1.3$  &   \cite{1998ApJ...496..196U}    \\ 
\object{NGC 4203}   & * &        &   $+0.38$  &   \cite{2001ApJ...562L.133U}    \\ 
\object{NGC 4579}   & * &        &   $+0.20$  &   \cite{2001ApJ...562L.133U}    \\ 
\object{NGC 5506}   & * &   B0   &   $+0.06$  &   This paper                    \\ 
                    &   &   B1   &   $-0.73$  &   This paper                    \\ 
                    &   &   B2   &   $-0.30$  &   This paper                    \\ 
\object{NGC 5793}   & * &   C1C  &   $-0.70$  &   \cite{2000AaA...360...49H}    \\ 
                    &   &   C1NE &   $-1.00$  &   \cite{2000AaA...360...49H}    \\ 
                    &   &   C2C  &   $-0.46$  &   \cite{2000AaA...360...49H}    \\ 
                    &   &   C2W  &   $-0.72$  &   \cite{2000AaA...360...49H}    \\ 
                    &   &   C2E  &   $+0.13$  &   \cite{2000AaA...360...49H}    \\ 
\object{NGC 7469}   & * &   2    &   $-0.34$  &   Norris et al., unpublished    \\ 
                    &   &   3    &   $+0.09$  &   Norris et al., unpublished    \\ 
                    &   &   5    &   $-0.12$  &   Norris et al., unpublished    \\ 
\object{NGC 7674}   &   &        &   $-1.78$  &   This paper                    \\ 
\hline
\end{tabular}
\caption{Seyferts from Table~\ref{tab:seyferts} that have
dual-frequency (18~cm and 6~cm) VLBI observations with matching beam
sizes, from which the spectral index is derived. Asterisks denote
objects with at least one component with a flat or inverted spectrum.}
\label{tab:spectra}
\end{center}
\end{table*}

\begin{table*}[ht!]
\begin{center}
\begin{tabular}{llrrrr}
\hline
\hline
Sample & &
\multicolumn{1}{c}{$N_{\alpha>-0.3}$} &
\multicolumn{1}{c}{$N_{\rm total}$} &
\multicolumn{1}{c}{\%} &
\multicolumn{1}{c}{$P$}\\
\hline
This paper                 & our obs + lit search  &  17 &  38 &  45\\
\cite{1987MNRAS.228..521U} & X-ray Seyferts        &   1 &  20 &   5 &  $  0.3$\,\%\\
\cite{1996ApJ...473..130R} & CfA Seyferts          &   3 &  46 & 6.5 &  $< 0.1$\,\%\\
\cite{1996ApJ...473..130R} & 12 $\mu$m Seyferts    &   4 &  40 &  10 &  $< 0.1$\,\%\\
\cite{1999AAS..137..457M}  & Distance limited      &  13 &  60 &  22 &  $  1.8$\,\%\\
\cite{2001ApJS..133...77H} & Palomar Seyferts      &  21 &  45 &  47 &  $  50$\,\%\\
\cite{1984ApJ...285..439U} & Distance limited      &   3 &  40 & 7.5 &  $< 0.1$\,\%\\
\hline
\end{tabular}
\caption{Fraction of flat- and inverted-spectrum components in various
Seyfert samples.  The column '$N_{\alpha>-0.3}$' gives the number of
components with spectral indices $\alpha>-0.3$ (i.e., components with
flat or inverted spectra).  The column '$N_{\rm total}$' gives the
total number of components in each sample.  The column \% gives
$(N_{\alpha>-0.3}/N_{\rm total})\times100$.  The column $P$ gives the
level of significance from a comparison of the fraction of flat- and
inverted-spectrum components in the VLBI sample of this paper (limits
treated as detections), to that of each comparison sample at arcsec
resolution, using the difference-of-two-proportions test with Yates
correction for continuity (e.g. \citealt{Glanz1992}).}
\label{tab:spectra-stats}
\end{center}
\end{table*}

\begin{table*}[ht!]
\begin{center} 
\begin{tabular}{llrrrrr}
\hline 
\hline
&&\multicolumn{5}{c}{Level of Significance}\\
Sample & &
\multicolumn{1}{c}{$\chi^2$} &
\multicolumn{1}{c}{KS} &
\multicolumn{1}{c}{Logrank} &
\multicolumn{1}{c}{Gehan} &
\multicolumn{1}{c}{Peto-Peto}\\
\hline
\cite{1987MNRAS.228..521U} & X-ray Seyferts      & 0.3\,\%  & 0.7\,\% &  0.3\,\% & 8.3\,\% & 8.3\,\%\\
\cite{1996ApJ...473..130R} & CfA Seyferts        & 0.1\,\%  & 0.1\,\% &  0.0\,\% & 0.5\,\% & 0.5\,\%\\
\cite{1996ApJ...473..130R} & 12\,$\mu$m Seyferts & 0.1\,\%  & 0.6\,\% &  0.0\,\% & 2.0\,\% & 2.0\,\%\\
\cite{1999AAS..137..457M}  & Distance Limited    & 0.1\,\%  & 0.3\,\% &  0.0\,\% & 1.2\,\% & 1.2\,\%\\
\cite{2001ApJS..133...77H} & Palomar Seyferts    & 1.0\,\%  &  27\,\% &   44\,\% &  75\,\% &  77\,\%\\
\cite{1984ApJ...285..439U} & Distance Limited    & 0.1\,\%  & 0.3\,\% &  0.0\,\% & 0.7\,\% & 0.7\,\%\\
\hline
\end{tabular}
\caption{Comparison of the spectral index distribution of the VLBI
Seyfert sample to various other Seyfert samples.  The level of
significance gives the probability that the spectral index
distribution of the comparison sample and the VLBI seyfert sample in
this paper were drawn from the same parent population.}
\label{tab:spectra-stats2}
\end{center}
\end{table*}

\begin{table*}
\begin{center}
\begin{tabular}{lrrrrllr}
\hline
\hline
\multicolumn{1}{c}{Name} 
& \multicolumn{1}{c}{Res.}
& \multicolumn{1}{c}{$\Theta$ (pc)}
& \multicolumn{1}{c}{$\Theta$ (kpc)}
& \multicolumn{1}{c}{$\Delta\Theta$}
& \multicolumn{1}{c}{Array}
& \multicolumn{1}{c}{Refs.}
& \multicolumn{1}{c}{$\Theta$ (opt.)}\\
& \multicolumn{1}{c}{(pc)}
& \multicolumn{1}{c}{(deg)}
& \multicolumn{1}{c}{(deg)}
& \multicolumn{1}{c}{(deg)}
& 
& 
& \multicolumn{1}{c}{deg}\\
\hline
\object{3C 390.3}   & 1.09      & 323  &     &   0  & global VLBI 5 GHz             & \cite{1996AaA...308..376A} &          \\
                    & 1210      &      & 323 &      & VLA 5 GHz                     & \cite{1996AaA...308..376A} &          \\
\object{Mrk 78}     & 1.75      & 40   &     &  55  & global VLBI 5 GHz             & \cite{VirLal2001}          &          \\
                    & 504       &      & 275 &      & VLA 5 GHz                     & \cite{VirLal2001}          &          \\
\object{Mrk 231}    & 0.343     & 93   &     & 77   & VLBA 15 GHz                   & \cite{1999ApJ...517L..81U} & 100      \\
                    & 3270      &      & 170 &      & VLA 1.5/8.4/15 GHz            & \cite{1999ApJ...516..127U} &          \\
\object{Mrk 1218}   & 0.582     & 330  &     &  13  & global VLBI 5 GHz             & \cite{VirLal2001}          & 130$^b$  \\
                    & 294       &      & 317 &      & VLA 5 GHz                     & \cite{VirLal2001}          &          \\
\object{NGC 1052}   & 0.015     &  65  &     &  30  & VLBA 5/8.4/22/43 GHz          & \cite{Kadler2002}          &          \\
                    & 162       &      & 275 &      & MERLIN 1.4 GHz                & \cite{Kadler2003}          &          \\
\object{NGC 1068}   & 0.221     &  11  &     &  29  & VLBA 1.7,5,15 GHz             & \cite{1998ApJ...504..147R} & 160      \\
                    & 217       &      & 40  &      & VLA 5 GHz                     & \cite{2001ApJS..133...77H} &          \\
\object{NGC 1167}   & 0.895     & 140  & 313 &   7  & global VLBI 5 GHz             & \cite{2001ApJ...552..508G} &          \\
                    & 131       &      & 66  &  74  & VLA 5 GHz                     & \cite{1995AaA...295..629S} &          \\
\object{NGC 1275}   & 0.416     & 353  &     &  23  & VLBA 2.3/5/8.4/15.4/22/43 GHz & \cite{2000ApJ...530..233W} &          \\
                    & 382       &      & 150 &      & VLA 1.4 GHz                   & \cite{1990MNRAS.246..477P} &          \\
\object{NGC 3079}   & 0.044     & 125  &     &  65  & VLBA 5/8/22 GHz               & \cite{1998ApJ...495..740T} &  75      \\
                    & 310       &      &  60 &      & VLA 1.4/4.9                   & \cite{1983ApJ...273L..11D} &          \\
\object{NGC 4151}   & 0.116     & 257  &     &   3  & VLBA 1.6,5 GHz                & \cite{1998ApJ...496..196U} & 140      \\
                    & 104       &      &  80 &      & VLA 5 GHz                     & \cite{2001ApJS..133...77H} &          \\
\object{NGC 4258}   & 0.029     &   3  &     &   8  & VLBA 22 GHz                   & \cite{1997ApJ...475L..17H} &  60      \\
                    & 125       &      & 355 &      & VLA 1.5 GHz                   & \cite{2000ApJ...536..675C} &          \\
\object{NGC 7212}   & 1.37      & 257  &     &  52  & global VLBI 5 GHz             & \cite{VirLal2001}          & 133$^a$  \\
                    & 609       &      &  25 &      & VLA 5 GHz                     & \cite{VirLal2001}          &          \\
\object{NGC 7469}   & 0.844     &  28  &     & 35   & global VLBI 5 GHz             & \cite{VirLal2001}          &  35      \\
                    & 193       &      & 353 &      & VLA 5 GHz                     & \cite{VirLal2001}          &          \\
\object{NGC 7674}   & 2.80      & 274  &     &  22  & VLBA+Y27+Arecibo 1.4 GHz      & \cite{Momjian2003}         &  64$^a$  \\
                    & 178       &      & 296 &      & MERLIN 1.7                    & \cite{Momjian2003}         &          \\
\hline
\object{III Zw 2}   & 0.260     & 303  &     &  63  & VLBA 15/43 GHz                & \cite{2000AaA...357L..45B} &          \\
                    & 3460      &      & 240 &      & VLA 1.5 GHz                   & Brunthaler, priv. comm.    &          \\
\object{Ark 564}    & 0.598     & 282  &     &      & global VLBI 5 GHz             & \cite{VirLal2001}          &  28      \\
\object{IC 5063}    & 3.30      &  66  &     &  49  & LBA 2.3 GHz + LBA HI          & \cite{2000AJ....119.2085O} &          \\
                    & 242       &      & 115 &      & ATCA 8.3/1.4GHz               & \cite{1998AJ....115..915M} &          \\
\object{NGC 2110}   & 0.151     & 350  &     &  16  & VLBA 8.4 GHz                  & \cite{2000ApJ...529..816M} &          \\
                    & 240       &      & 186 &      & VLA 1.4/5 GHz                 & \cite{1983ApJ...264L...7U} &          \\
\object{NGC 3147}   & 0.219     & 317  &     &      & VLBA 1.6/2.3/5/8.4 GHz        & \cite{2001ApJ...562L.133U} &  65      \\
\object{NGC 5506}   & 0.815     &  73  &     &  76  & EVN 1.6/5 GHz                 &      This paper            &   1      \\
                    & 359       &      & 177 &      & VLA 4.9 GHz                   & \cite{1996ApJ...467..551C} &          \\
\object{NGC 7682}   & 0.996     & 125  &     &      & global VLBI 5 GHz             & \cite{VirLal2001}          &  68$^b$  \\
\hline
\end{tabular}
\caption{Subset of sources from Table~\ref{tab:seyferts} for which both
pc and kpc-scale structures have been observed. The misalignment
angle, $\Delta\Theta$, is calculated as $\Theta({\rm pc})-\Theta({\rm
kpc})$, and has then been reflected onto the range of $0^{\circ}$ to
$90^{\circ}$. For the seven objects at the end of the table, either
the P.A. information of the kpc scale structure was not completely
reliable or was unavailable.  The position angles of the galaxy minor
axes in the last column were drawn from \cite{1991trcb.book.....D},
except for those marked $a$ taken from \cite{Schmitt2000} and those
marked $b$ measured by us from Hubble Space Telescope archival images
(see \S\ref{sec:misalignment}).}
\label{tab:misalignment}
\end{center}
\end{table*}

\end{document}